\documentclass[11pt]{article}

\usepackage{jcappub} 

\usepackage[T1]{fontenc} 

\usepackage{amssymb}
\usepackage{graphicx, epsfig, bm, amsmath,mathtools}

\usepackage{hyperref}
\usepackage{cleveref}

\usepackage{subcaption}
\usepackage{soul}

\newcommand{\ud}{\mathrm{d}}
\newcommand{\dd}[2]{\frac{\ud {#1} }{\ud {#2}}}

\newcommand{\pd}[2]{\frac{\partial {#1} }{\partial {#2}}}

\DeclareMathOperator{\diag}{diag}

\newcommand{\Mpl}{M_\mathrm{Pl}}

\newcommand{\gambar}{\bar{\gamma}_{ij}}

\newcommand{\Aij}{\tilde{A}_{ij}}
\newcommand{\Aiju}{\tilde{A}^{ij}}

\usepackage{color}
\usepackage{ifthen}

\bibstyle{JHEP}

\title{Gauge preheating with full general relativity}

\author[a]{Peter Adshead,}
\author[b,c,d]{John T. Giblin, Jr.,}
\author[e]{Ryn Grutkoski,}
\author[f]{and Zachary J. Weiner}

\affiliation[a]{Illinois Center for Advanced Studies of the Universe \& Department of Physics, University of Illinois at Urbana-Champaign, Urbana, IL 61801, U.S.A.}
\affiliation[b]{Department of Physics, Kenyon College, Gambier, Ohio 43022, U.S.A.}
\affiliation[c]{Department of Physics/CERCA/Institute for the Science of Origins, Case Western Reserve University, Cleveland, OH 44106, U.S.A.}
\affiliation[d]{Center for Cosmology and AstroParticle Physics (CCAPP) and Department of Physics, Ohio State University, Columbus, OH 43210, U.S.A.}
\affiliation[e]{Department of Astronomy \& Astrophysics, University of Chicago, Chicago, IL 60637, U.S.A.}
\affiliation[f]{Department of Physics, University of Washington, Seattle, WA 98195, U.S.A.}
\emailAdd{adshead@illinois.edu}
\emailAdd{giblinj@kenyon.edu}
\emailAdd{rmgrutkoski@uchicago.edu}
\emailAdd{zweiner@uw.edu}

\abstract{ We study gauge preheating following pseudoscalar-driven inflation in full general relativity. We implement the Baumgarte-Shapiro-Shibata-Nakamura (BSSN) scheme to solve the full nonlinear evolution of the metric alongside the dynamics of the pseudoscalar and gauge fields. The dynamics of the background and emission of gravitational waves are broadly consistent with simulations in a Friedmann-Lema\^{i}tre-Robertson-Walker (FLRW) spacetime.
We find large, localized overdensities in the BSSN simulations of order $\delta = \delta\rho/\rho \sim 30$, and the dimensionless power spectrum of $\delta$ peaks above unity. These overdense regions are seeded on length scales only slightly smaller than the horizon, and  have a compactness $C \sim 0.1$. The scale of peak compactness is shorter than the Jeans length, which implies that pressure of the matter fields plays an important role in the evolution of these objects.
\\
{\bf Date:} \today }

\begin{document}
\maketitle
\flushbottom


\section{Introduction}
\label{sec:intro}

The end of cosmic inflation~\cite{Guth:1980zm, Linde:1981mu, Albrecht:1982wi, Linde:1983gd} and the subsequent transition to the radiation-dominated hot Big Bang remains one of the most poorly understood epochs in the evolution of the Universe. During this reheating epoch, the accelerated expansion of inflation must end, and the inflaton energy density must be eventually transferred to relativistic degrees of freedom to begin the hot Big Bang.  Because the scales involved are so small---smaller than the Hubble radius at the end of inflation---information about the reheating epoch is either erased as the resulting Standard Model plasma achieves thermal equilibrium or is inseparable from the effects of the subsequent nonlinear gravitational evolution of structure formation.

While reheating may be facilitated by perturbative decays of the inflaton to other particles~\cite{Abbott:1982hn, Albrecht:1982mp}, the homogeneous, oscillating inflaton background can source explosive production of particles and rapid growth of matter inhomogeneities via preheating~\cite{Traschen:1990sw,Shtanov:1994ce,Kofman:1994rk, Kofman:1997yn}.
The collective dynamics of the oscillating background has long been a fertile ground for model building and searches for observable signatures of reheating~\cite{Greene:1997fu, Kaiser:1997mp, Parry:1998pn, Bassett:1998wg, Easther:1999ws, Bassett:1999mt, Felder:2006cc, Easther:2010mr} (for a review, see  Ref.~\cite{Amin:2014eta}).
Possible gravitational relics, such as gravitational waves~\cite{Khlebnikov:1997di, Easther:2006gt, Easther:2006vd, Garcia-Bellido:2007nns, Easther:2007vj, Dufaux:2007pt, Dufaux:2008dn, Dufaux:2010cf, Cosme:2022htl}, collapsed structures like primordial black holes~\cite{Bassett:1998wg, Green:2000he, Jedamzik:2010dq, Martin:2019nuw, Musoke:2019ima, Auclair:2020csm, Martin:2020fgl, Eggemeier:2021smj}, or compact mini halos~\cite{Bringmann:2011ut, Aslanyan:2015hmi}, offer potential probes into the reheating epoch.

Preheating into gauge fields---gauge preheating---is an extremely violent process. In models of gauge preheating, effectively all of the energy stored in the inflaton can be transferred into gauge field radiation within a single oscillation of the inflaton about the minima of its potential~\cite{Armendariz-Picon:2007gbe, Deskins:2013dwa, Adshead:2015pva, Adshead:2016iae, Cheng:2015oqa, Adshead:2017xll, Cuissa:2018oiw, Figueroa:2023oxc}. This rapid energy transfer is facilitated by a tachyonic instability in the gauge field sourced by the rolling inflaton field.
The tachyonic enhancement of gauge fields by rolling pseudoscalars (a long-appreciated phenomenon~\cite{Carroll:1991zs, Garretson:1992vt, Prokopec:2001nc}) can also realize strong backreaction on the homogeneous motion of the pseudoscalar inflaton \cite{Freese:1990rb, Adams:1992bn}, enabling models of inflation on steep potentials\footnote{Recent work has further highlighted an instability in the strong backreaction regime of axion inflation coupled to gauge fields~\cite{Domcke:2020zez, Gorbar:2021rlt, Peloso:2022ovc, Domcke:2023tnn, Garcia-Bellido:2023ser, vonEckardstein:2023gwk}, but without nontrivial, \textit{ad hoc} model constructions such a scenario is wholly precluded.}~\cite{Anber:2009ua, Adshead:2012kp} and warm inflation~\cite{Berghaus:2019whh}.
Away from the regime of strong backreaction, perturbative backreaction of the gauge modes during and after inflation may  produce observable non-Gaussianity~\cite{Barnaby:2010vf,Barnaby:2011vw,Barnaby:2011qe, Barnaby:2012tk, Anber:2012du}, chiral gravitational waves~\cite{Sorbo:2011rz, Cook:2011hg, Sorbo:2011rz, Anber:2012du, Martinec:2012bv,  Adshead:2013qp, Adshead:2013nka,  Maleknejad:2012fw,  Namba:2015gja, Adshead:2016omu, Caldwell:2017chz, Weiner:2020sxn}, primordial black holes~\cite{Linde:2012bt, Bugaev:2013fya}, and primordial magnetic fields~\cite{Garretson:1992vt,Prokopec:2001nc, Barnaby:2012tk, Fujita:2015iga, Adshead:2016iae} and  the baryon asymmetry~\cite{Alexander:2004us, Maleknejad:2012wqk, Anber:2015yca, Kamada:2016eeb, Maleknejad:2016dci, Caldwell:2017chz, Adshead:2017znw, Domcke:2019mnd, Domcke:2022kfs}.

The strongest effects of gauge field production occur when the inflaton rolls the fastest, which typically occurs at the end of inflation.
The large couplings required to produce effects observable on scales accessible to the cosmic microwave background (CMB) or gravitational-wave interferometers subsequently generate a high-frequency gravitational wave background during preheating so large that existing bounds on the effective number of relativistic species rules the regime out~\cite{Adshead:2018doq, Adshead:2019lbr, Adshead:2019igv}.\footnote{To avoid these effects, rolling spectator fields have been instead invoked to source effects on observable scales today. These fields source transient effects and avoid spoiling the inflationary solution and overproducing gravitational waves at the end of inflation~\cite{Cook:2013xea, Namba:2015gja, Maleknejad:2016qjz, Ozsoy:2020ccy,Ozsoy:2021onx, Campeti:2022acx,Fujita:2023inz, Unal:2023srk}, but cannot realize reheating after inflation.}  The production of large metric perturbations calls into question whether nonlinear gravitational effects can be safely neglected. Since gravitational wave backgrounds from gauge preheating currently provide the strongest constraint on these  models, it is crucial to test the robustness of prior predictions~\cite{Adshead:2018doq, Adshead:2019igv, Adshead:2019lbr} and whether nonlinear gravity might enhance \cite{Bastero-Gil:2010tpb} or suppress \cite{Kou:2021bij} the production of gravitational waves. Further, the production of large metric fluctuations indicates the presence of large inhomogeneities in the matter sector which may undergo gravitational collapse under the influence of local gravity. The purpose of this work is to push further the exploration of gravitational signatures of preheating into the regime of local,  nonlinear gravity.  To that end, building on the results of \cite{Giblin:2019nuv}, we initiate a study of preheating into gauge fields using the full machinery of numerical relativity \cite{Shibata:1995we,Baumgarte:1998te,Baumgarte:2010ndz} to follow the nonlinear evolution of the metric.

In this paper, we demonstrate that preheating probes regions where nonlinear gravitational effects are expected to become important, and we explore the degree to which simulations that include nonlinear gravity are required to accurately characterize this epoch. In particular, we perform a careful study of  gravitational effects during gauge preheating, focusing on the development of large density inhomogeneities and gravitational wave production. We first demonstrate that linearized gravity is quickly violated. Then, using numerical relativity, we follow the evolution of the metric including the effects of dynamical gravity to show that this breakdown of linearized gravity does not signal the formation of black holes. We demonstrate that, despite producing regions with very large density contrasts $\delta\rho/\rho\sim 30$, there is nevertheless no evidence that black holes are formed---no horizons are formed in our simulations. We show that  the spatial extent of the regions of large density contrast are smaller than the Jeans length, which indicates that pressure plays an important role in their subsequent evolution. Furthermore,  their compactness, which is a measure of whether a region satisfies the so-called `hoop-conjecture' \cite{Misner:1973prb}, attains a maximum value of $C \sim 10^{-1}$.

The paper is structured as follows. In \cref{sec:background} we define the model and present the equations that govern the evolution and amplification of gauge fields during and after inflation.
In section \ref{sec:results} we present results of numerical simulations, focusing on the dynamics of density and gravitational wave fluctuations as a function of the axion-gauge coupling strength. Our conclusions and proposed avenues for future work are presented in \cref{sec:conclusions}. The details of the decomposition of gauge fields in the BSSN formalism are relegated to appendix \ref{app:BSSNdeets}, while appendix \ref{app:perttheory} details how our initial conditions are set in perturbation theory in the BSSN formalism, and finally appendix \ref{app:robustness} describes robustness checks of our results.

We use natural units, which set $\hbar = c = 1$, and define the reduced Planck mass
$\Mpl = 1/\sqrt{8\pi G}$. Repeated/contracted Greek spacetime indices are summed via the Einstein summation convention.


\section{Gauge preheating and the BSSN formalism}\label{sec:background}

We consider a pseudoscalar inflaton, $\varphi$, minimally coupled to Einstein gravity, and coupled to a U(1) gauge field, $A_\mu$, described by the action
\begin{align}\label{eqn:lagrangian}
    S &= \int \ud^4 x \, \sqrt{-g} \left[
        \frac{\Mpl^2}{2} R- \frac{1}{2} \nabla_\mu \varphi \nabla^\mu \varphi
        - V(\varphi)
        - \frac{1}{4} F_{\mu\nu} F^{\mu\nu}
        - \frac{X(\varphi)}{4} F_{\mu\nu} \tilde{F}^{\mu\nu}\right].
\end{align}
where $R$ is the Ricci scalar and $F_{\mu\nu} =  \nabla_\mu A_\nu - \nabla_\nu A_\mu$ is the field strength tensor whose dual is
\begin{align}
    \tilde{F}^{\mu\nu} = \frac{1}{2} \epsilon^{\mu\nu\alpha\beta} F_{\alpha\beta}.
\end{align}
The Levi-Civita tensor is
\begin{equation}
    \epsilon_{\mu_1 \mu_2 \cdots \mu_n} = \sqrt{- g} \varepsilon_{\mu_1 \mu_2 \cdots \mu_n},
\end{equation}
which is written in terms of the permutation symbol (the Levi-Civita symbol), $\varepsilon_{\mu_1 \mu_2 \cdots \mu_n}$.
We work with the convention that $\epsilon^{\mu_1 \mu_2 \cdots \mu_n} \equiv \mathrm{sign}\, g \, \varepsilon^{\mu_1 \mu_2 \cdots \mu_n} / \sqrt{- g}$, such that
$\varepsilon^{\mu_1 \mu_2 \cdots \mu_n} \equiv \varepsilon_{\mu_1 \mu_2 \cdots \mu_n}$.
Here $\nabla_\mu$ denotes the ($\mu$ component of) the four dimensional covariant derivative compatible with the full metric $g_{\mu \nu}$.

We consider a simple toy model of inflation specified by the potential~\cite{Linde:1981mu}
\begin{align}
    V = \frac{1}{2} m^2 \varphi^2.
\end{align}
Planck's best-fit scalar spectral amplitude
$A_s \approx 2.1 \times 10^{-9}$~\cite{Planck:2018vyg} sets $m = 6.05 \times 10^{-6} \, \Mpl$.
Though inflation driven by a monomial potential is strongly disfavored~\cite{Planck:2018jri,BICEP:2021xfz}, we choose a quadratic potential
merely to provide a simple model for the inflaton's dynamics during preheating. For further discussion of the effects of potential choice on gauge preheating, see~\cite{Adshead:2019lbr}.
We consider the standard shift-symmetric, dimension-5 axial coupling between the pseudoscalar and the  gauge field \begin{align}
    X(\varphi)
    &= \frac{\alpha_g}{\Mpl} \varphi.
\end{align}

Numerical investigations of preheating typically employ a homogeneously expanding, Friedmann-Lema\^{i}tre-Robertson-Walker (FLRW)
background spacetime with metric
\begin{align}
    \label{eqn:flrw-metric}
    g_{\mu\nu}^{\rm FLRW}
    &= a(\tau)^2 \eta_{\mu\nu} = a(\tau)^2 \diag\left[ -1, 1, 1,1 \right].
\end{align}
Here $\tau$ is conformal time, and the scale factor $a(\tau)$ evolves according to the Friedmann equations
\begin{equation}
    \mathcal{H}^2
    \equiv \left(\frac{{a^\prime}}{a}\right)^2
    = \frac{a^2}{3 \Mpl^2} \bar{\rho}\, ,
\end{equation}
and
\begin{equation}
    \mathcal{H}'
    + \mathcal{H}^2
    = \frac{a^2}{6 \Mpl^2} \left( \bar{\rho} - 3 \bar{P} \right).
\end{equation}
Primes denote a derivative with respect to conformal time, $\tau$, and overbars generally indicate averaged quantities on constant-$\tau$ hypersurfaces.
The (averaged) energy density and pressure are
$\bar{\rho} \equiv - \bar{T}^0_{\hphantom{0} 0}$ and $\bar{P} \equiv \delta_i^{\hphantom{i} j} \bar{T}^i_{\hphantom{i}j} / 3$, respectively.

In an FLRW spacetime, the equations of motion for the scalar field and gauge fields are
\begin{subequations}\label{eqn:flrw-eles}
\begin{align}
    \varphi'' + 2 \mathcal{H} \varphi' - \partial_i \partial_i \varphi
        + a^2 \left(
            \dd{V}{\varphi}
            + \frac{1}{4} \dd{X}{\varphi} F_{\mu \nu} \tilde{F}^{\mu \nu}
        \right)
    &= 0,
    \label{eqn:varphi-eom-flrw}
    \\
    \partial_i A_i'
        - \partial_i \partial_i A_{0}
        - \epsilon^{i j k} \partial_i X(\varphi) \partial_j A_k
    &= 0,
    \label{eqn:gauss-law-flrw}
    \\
    A_i''
        - \partial_j \partial_j A_i
        - \partial_i \left(A_0' - \partial_j A_j \right)
        - \partial_\mu X(\varphi) \frac{1}{2} \varepsilon^{\mu i \rho\sigma} F_{\rho\sigma}
    &= 0.
    \label{eqn:Ai-eom-flrw}
\end{align}
\end{subequations}
In these equations, repeated Latin (i.e., spatial) indices are implicitly contracted with the
Kronecker delta function.
Fixing the (flat-space) Lorenz gauge $\eta^{\mu \nu} \partial_\mu A_\nu = 0$, \cref{eqn:gauss-law-flrw} and \cref{eqn:Ai-eom-flrw} may both be recast into the second-order
differential equations taking the form
\begin{align}
    A_\beta''
        - \partial_i \partial_i A_\beta
        - \partial_\mu X(\varphi) \eta_{\beta\nu}
            \frac{1}{2} \varepsilon^{\mu\nu\rho\sigma} F_{\rho\sigma}
     &= 0.
    \label{eqn:flat-lorenz-ele-abelian}
\end{align}


\subsection{Numerical Relativity} \label{numrel}

To implement nonlinear gravity, we use the BSSN decomposition~\cite{Baumgarte:1998te,Shibata:1995we} where the metric is decomposed as
\begin{align}
    g_{\mu \nu}^\mathrm{BSSN}
    &= \begin{pmatrix}
            - \alpha^2 + \beta_l \beta^l & \beta_j \\
            \beta_i & \gamma_{i j}
        \end{pmatrix}.
    \label{eqn:bssn-metric}
\end{align}
The lapse $\alpha$ and the shift $\beta^i$ parameterize (nondynamical) gauge degrees of freedom.
Three-dimensional hypersurfaces are measured by the spatial metric\footnote{Note that the overbar in this expression does not refer to a spatial average. In keeping with the notation in the BSSN community, an overbar here denotes the unit determinant part of the spatial metric.}
\begin{align}
    \gamma_{ij} = e^{4\phi} \bar{\gamma}_{ij},
\end{align}
which is further decomposed into a conformal factor $\phi$ and a unit-determinant spatial metric $\bar{\gamma}_{ij}$.
We denote spatial covariant derivatives (those compatible with the spatial metric $\gamma_{i j}$) with $D_i$, and indicate trace-removed quantities as
\begin{align}
X_{i j}^\mathrm{TF} \equiv X_{i j} - \frac{1}{3}\gamma_{i j} \gamma^{m n} X_{m n} .
\end{align}
Three-dimensional hypersurfaces are defined relative to the temporal coordinate $t$ with normal vector
\begin{align}\label{eqn:normal-vector-def}
    n^\mu = \frac{1}{\alpha} \left(1, - \beta^i \right).
\end{align}

The evolution of the metric components is specified by the set of first-order differential equations,
\begin{subequations}\label{eqn:bssn-metric-eoms}
\begin{align}
    \partial_t \phi
    &= - \frac{1}{6} \alpha K
        + \beta^i \partial_i \phi
        + \frac{1}{6}\partial_i \beta^i,
    \\
    \partial_t \bar{\gamma}_{ij}
    &= -2 \alpha \Aij
        + \beta^k \partial_k \gambar
        + \bar{\gamma}_{ik} \partial_j \beta^k
        + \bar{\gamma}_{kj} \partial_i \beta^k
        - \frac{2}{3} \bar{\gamma}_{ij} \partial_k \beta^k,
    \\
    \partial_t K
    &= \gamma^{ij} D_j D_i \alpha
        + \alpha \left( \Aij \Aiju + \frac{1}{3} K^2 \right)
        + \frac{\alpha}{2 \Mpl^2} \left( \rho + S \right)
        + \beta^i \partial_i K,
    \\
    \begin{split}
        \partial_t \Aij
        &= e^{-4\phi} \left[
            -  D_j D_i \alpha
            + \alpha \left( R_{ij} - S_{ij} / \Mpl^2 \right)
        \right]^\mathrm{TF}
        + \alpha \left( K \Aij -2 \tilde{A}_{il} \tilde{A}^l_j \right)
        + \beta^k \partial_k \Aij
        \\ &\hphantom{{}={}}
        + \tilde{A}_{ik} \partial_j \beta^k + \tilde{A}_{kj}\partial_i \beta^k -\frac{2}{3}\Aij \partial_k \beta^k,
    \end{split}
\end{align}
\end{subequations}
where $D_i$ is the 3-dimensional covariant derivative, $K$ is the trace of the extrinsic curvature tensor, $\Aij$ is the traceless part of the extrinsic curvature, see eqs.\ \eqref{eqn:def-extrinsic-curvature} and \eqref{eqn:trace-extrinsic-curvature-div-n}, and $R_{ij}$ is the spatial projection of the Ricci tensor. The sources, $\rho$, $S$, and $S_{ij}$,  are calculated from the stress-energy tensor, see \cref{3p1decomp} for details.
The BSSN system introduces new degrees of freedom so that Einstein's equations are computationally stable.  Numerical solutions must satisfy the Hamiltonian and momentum constraints,
\begin{equation}
\mathcal{H}=\bar{\gamma}^{ij} \bar{D}_i \bar{D}_j e^{\phi} -\frac{e^{\phi}}{8}\bar{R} +\frac{e^{5\phi}}{8}\tilde{A}_{ij}\tilde{A}^{ij} - \frac{e^{5\phi}}{12} K^2 +2\pi e^{5\phi} \rho =0,
 \label{eq:hamconst}
\end{equation}
and
\begin{equation}
\mathcal{M}^i=\bar{D}_j (e^{6\phi} \tilde{A}^{ji} ) - \frac{2}{3}e^{6 \phi} \bar{D}^i K -8\pi e^{10\phi} S^i = 0,
 \label{eq:momconst}
\end{equation}
to sufficient precision throughout the simulations.\footnote{While the conformal Hubble scale and the Hamiltonian constraint use the same symbol, $\mathcal{H}$, they both are standard.  Throughout the text we will specify which quantity is being considered.}

Since the lapse $\alpha$ and shift $\beta^i$ are purely gauge degrees of freedom, we are free to specify them via convenient evolution equations.  To allow for the formation of compact structures, such as black holes, while also trying to stay near the FLRW background~\cite{Giblin:2018ndw} we take a {\sl Bona-Mass\'o} slicing condition for the lapse~\cite{Bona:1994dr,Baumgarte:2010ndz},
\begin{equation}
   \left( \partial_t  - \beta^i\partial_i\right)\alpha = - \frac{\alpha^2}{3} K,
    \label{eq:lapse}
\end{equation}
as well as a hyperbolic gamma driver condition for the shift,
\begin{align}
\partial_t \beta^i &= \frac{3}{4}B^i, \\
\partial_t B^i &= \partial_t \bar{\Gamma}^i-\frac{\eta}{2}B^i,
\label{eq:shift}
\end{align}
where we set $\eta = 100$ as a choice that minimizes the violation of the constraints \cref{eq:hamconst,eq:momconst} at late times (see \cref{app:robustness}).  In the homogeneous limit, this slicing reduces to
\begin{equation}
    \partial_t \alpha = -\frac{\alpha^2 K}{3} = \alpha^2 H,
\end{equation}
which is the normal conformal-time evolution equation for the scale factor when $\alpha \rightarrow a$.

We track the scalar field $\varphi$ and its conjugate momentum, $\Pi$, which is the component of its
(covariant) four-gradient normal to spatial hypersurfaces,
\begin{align}\label{eqn:Pi-def}
    \Pi
    &\equiv n^\mu \nabla_\mu \varphi
    = \frac{1}{\alpha} \left(
        \nabla_0 \varphi - \beta^k \nabla_k \varphi
    \right).
\end{align}
We additionally promote the spatial derivatives of the scalar field to dynamical quantities,
\begin{align}
\psi_i \equiv D_i \varphi,
\end{align}
and split the vector potential $A_\mu$ into its components along and orthogonal to spatial
hypersurfaces,
\begin{equation}
    \mathcal{A} = - n^\nu A_\nu,
\end{equation}
and
\begin{equation}
    \mathcal{A}_\mu = \gamma_\mu^{\hphantom{\mu}\nu} A_\nu,
\end{equation}
respectively, such that standard four-potential is simply reconstructed as
$A_\mu \equiv \mathcal{A}_\mu + n_\mu \mathcal{A}$, see \cref{appsec:GFEOMS}.
The electric and magnetic fields are then given by
\begin{align}
    E^\mu
    &= \gamma^\mu_{\hphantom{\mu} \nu} n_\alpha F^{\nu \alpha},
    \label{eqn:def-E-mu}
    \\
    B^\mu
    &= - \gamma^\mu_{\hphantom{\mu} \nu} n_\alpha \tilde{F}^{\nu \alpha}
    \label{eqn:def-B-mu}.
\end{align}
In practice, we can fully evolve the system by evolving
$\mathcal{A}$, $\mathcal{A}_m$, and the purely spatial vector $E^m$.
In terms of these variables, the scalar field's Euler-Lagrange equation, \cref{eqn:scalar-ele}, reduces to the first-order system
\begin{subequations}\label{eqn:scalar-field-eom-3-plus-1}
\begin{align}
    \partial_t \varphi
    &= \beta^m D_m \varphi + \alpha \Pi,
    \\
    \partial_t \psi_m
    &= \beta^n \partial_n \psi_m
        + \psi_n \partial_m \beta^n
        + \alpha D_m \Pi
        + \Pi D_m \alpha,
    \\
\begin{split}
    \partial_t \Pi
    &= \beta^m D_m \Pi
        + e^{- 4 \phi} \bar{\gamma}^{m n} \left(
            \alpha \partial_m \psi_n
            + D_m \alpha \psi_n
        \right)
        \\ &\hphantom{{}={}}
        + \alpha \left(
            K \Pi
            - e^{- 4 \phi} \bar{\gamma}^{m n} \Gamma^o_{\hphantom{o} m n} \psi_o
        \right)
        + \alpha \left(
            - \dd{V}{\varphi}
            - \dd{X}{\varphi} E_m B^m
        \right).
\end{split}
\end{align} \label{eq:eomsphi}
\end{subequations}

In the BSSN system, the metric in  \cref{eqn:bssn-metric} is not conformally related to $\eta_{\mu \nu}$---unlike the FLRW metric \cref{eqn:flrw-metric}---and it is more convenient to chose the covariant Lorenz gauge
\begin{align}
\nabla^\mu A_\mu + Z = 0.
\end{align}
The auxiliary field $Z$ is a dynamical {\sl constraint-damping} degree of freedom that can increase the computational stability of the system~\cite{Palenzuela:2009hx, Hilditch:2013sba, Zilhao:2015tya, Clough:2022ygm}.
In practice, we do not find including $Z$ meaningfully improves the stability of our simulations,
but we retain it below for completeness.
With this choice, the equations of motion for the gauge field sector are
\begin{subequations}\label{eqn:gauge-eoms-3-plus-1}
\begin{align}
\begin{split}
    \partial_t E^m
    &= \beta^o \partial_o E^m
        - E^o \partial_o \beta^m
        + \epsilon^{m n o} D_n \alpha B_o
    \\ &\hphantom{{}={}}
        + \alpha \left(
            K E^m
            + \epsilon^{m n o} D_n B_o
            - \mathcal{J}^m
            + D^m Z
        \right),
        \label{eqn:electric-field-eom-standard-vars}
\end{split}
    \\
    \partial_t \mathcal{A}_m
    &= \beta^o \partial_o \mathcal{A}_m
        + \mathcal{A}_o \partial_m \beta^o
        - \alpha \left(
            E_m
            + D_m \mathcal{A}
        \right)
        - \mathcal{A} D_m \alpha,
    \\
    \partial_t \mathcal{A}
    &= \beta^o D_o \mathcal{A}
        + \alpha \left(
            K \mathcal{A}
            - D^m \mathcal{A}_m
            - Z
        \right)
        - \mathcal{A}^m D_m \alpha,
    \\
    \partial_t Z
    &= \beta^o D_o Z
        - \alpha \kappa Z
        + \alpha \left(
            D_m E^m
            - \mathcal{J}
        \right).
\end{align}
\end{subequations}
Here the ``source vector'' has components
\begin{subequations}\label{eqn:source-vector-3-plus-1}
\begin{align}
    \mathcal{J}
    &= X'(\varphi) B^m D_m \varphi,
    \\
    \mathcal{J}^m
    &= - X'(\varphi)
        \left(
            \Pi B^m
            + \epsilon^{m n o} D_n \varphi E_o
        \right).
\end{align}
\end{subequations}
Gauss's law requires the divergence of the electric field satisfy
\begin{align}
    \mathcal{G} = D_m E^m - \mathcal{J} = 0,
    \label{eq:GaussLaw}
\end{align}
which is an additional constraint on the system that must be satisfied throughout the simulation, see \cref{app:robustness}.  Finally the source-terms in  \cref{eqn:bssn-metric-eoms} evaluate to
\begin{subequations}
\begin{align}
    \rho
    &= \frac{1}{2} \Pi^2 + \frac{1}{2} D_m \varphi D^m \varphi + V(\varphi)
        + \frac{1}{2} \left( E_m E^m + B_m B^m \right),\label{eqn:rho}
    \\
    S_m
    &= - \Pi D_m \varphi
        + \epsilon_{m n o} E^n B^o,
    \\
\begin{split}
    S_{m n}
    &= D_m \varphi D_n \varphi
        -  \left(
            E_m E_n
            + B_m B_n
        \right)
    \\ &\hphantom{{}={}}
        - \gamma_{m n} \left[
            - \frac{1}{2} \Pi^2
            + \frac{1}{2} D_i \varphi D^i \varphi
            + V(\varphi)
            - \frac{1}{2}
            \left( E_o E^o + B_o B^o \right)
        \right],
\end{split}
    \\
    S
    &= \frac{3}{2} \Pi^2
        - \frac{1}{2} D_m \varphi D^m \varphi
        - 3 V(\varphi)
        + \frac{1}{2} \left( E_m E^m + B_m B^m \right),
\end{align}
\end{subequations}
which close the dynamical system.


\section{Results}
\label{sec:results}

In this section we present the results of our simulations. We begin in \cref{subsec:codeandback} by describing our software and the computational choices we make in our simulations.  We then compare the evolution of the background, including the full effects of nonlinear gravity, with our previous FLRW simulations. We then look for signs that gravity might lead to collapse in \cref{subsec:density}, and finally study the resulting gravitational wave spectra in \cref{subsec:GWs}.


\subsection{\textsc{GABERel}, initial conditions and background evolution}\label{subsec:codeandback}

We extend \textsc{GABERel}~\cite{Child:2013ria,Giblin:2019nuv} to treat gauge fields in addition to scalar fields (as presented in Ref.~\cite{Giblin:2019nuv}) in full numerical relativity using the BSSN formalism.
Such an analysis is the only way to assess whether nonlinear gravitational physics is important and whether it leads to the collapse of gravitationally bound objects.

The BSSN simulations we present here use grids with $N^3 = 384^3$ points and a comoving box length $L = 7.5 \, m^{-1}$, with $a=1$ at the end of inflation; therefore the initial, physical box size is $L_0 = e^{-2} L$ at the beginning of the simulation.
We use the standard fourth-order Runge-Kutta method for time integration with steps of size $\Delta t = \Delta x / 10 = L_0/(10N)$.
The spatial discretization uses fourth-order finite-difference stencils (with upwind variants for advective derivatives and centered differences otherwise).
We begin simulations two $e$-folds before the end of inflation, which ensures that even the largest-scale modes in the simulation begin with nearly Bunch-Davies initial conditions.
Each mode has a uniform-random phase and an amplitude sampled from the Rayleigh distribution with variance set by the Bunch-Davies vacuum.
The fields' time derivatives are set in the Wentzel-Kramers-Brillouin (WKB) approximation following the standard prescription~\cite{Felder:2000hq,Giblin:2019nuv}.
We filter out high-wavenumber modes from the initial conditions as in Ref.~\cite{Giblin:2019nuv}, setting the cutoff scale to $k_\star = 1/12 \cdot \pi \sqrt{3} / L$.

We can also compare the fully nonlinear system to the FLRW system from previous work.  To simulate the FLRW system, we use the same software described in Refs.~\cite{Adshead:2019igv,Adshead:2019lbr}, which also uses a fourth-order spatial discretization and time evolution scheme.
Because FLRW simulations are less computationally expensive, we use grids with $N^3 = 512^3$ points and a larger box length of $L = 15 \, m^{-1}$.
The initial conditions are cut off at a wavenumber $k_\star = 1/2 \cdot \pi \sqrt{3} / L$.  The FLRW simulations are therefore rather different from the BSSN ones, not just in the physical
content and its representation but also in numerical implementation.
The only quantitative comparisons that can be made between the two are statistical, for which reason the differing choices of simulation volumes and grids are relatively inconsequential.
Indeed, agreement between the two methods (in regimes where it is expected) provides a robust test of the results' independence of the numerical procedure.

The axion-gauge field coupling in \cref{eqn:varphi-eom-flrw} leads to a tachyonic instability in the gauge fields whenever $\dot{\phi}_0 \neq 0$ for momenta that satisfy~\cite{Carroll:1991zs, Prokopec:2001nc, Anber:2009ua}
\begin{align} \label{eq:GFinstabwindow}
aH < k < \alpha_g\frac{\dot{\phi}}{f}.
\end{align}
During inflation, this tachyonic instability leads to the exponential enhancement of one (helical) polarization of the gauge field relative to the other. Since the width of the instability in eq.\ \eqref{eq:GFinstabwindow}  is proportional to the axion velocity, the largest effects typically occur near the end of inflation where the inflaton velocity is the largest. These dynamics complicate the setting of initial conditions for preheating simulations.
As detailed in Refs.~\cite{Adshead:2019igv,Adshead:2019lbr}, the initial conditions are set by solving for the dynamics of the linearized equations of motion of the background and field fluctuations during inflation, again until two $e$-folds before inflation ends.  At this time, the initial conditions for gauge field fluctuations hardly depart from the Bunch-Davies vacuum on the scales present in the simulation (as noted above).  The dynamics of the homogeneous mode of the inflaton, however, are impacted by gauge-field backreaction; in both FLRW and BSSN simulations we set $\bar{\phi}_0$ and $\bar{\phi}_0'$ using the full numerical results.
We consider values of $\alpha_g$ between 8, the smallest for which preheating is efficient, and 14, which is roughly the largest value currently allowed by $\Delta N_\mathrm{eff}$ constraints on gravitational wave production derived in Refs.~\cite{Adshead:2018doq,Adshead:2019lbr,Adshead:2019igv}.

\begin{figure}[ht]
    \centering
    \includegraphics[width=\textwidth]{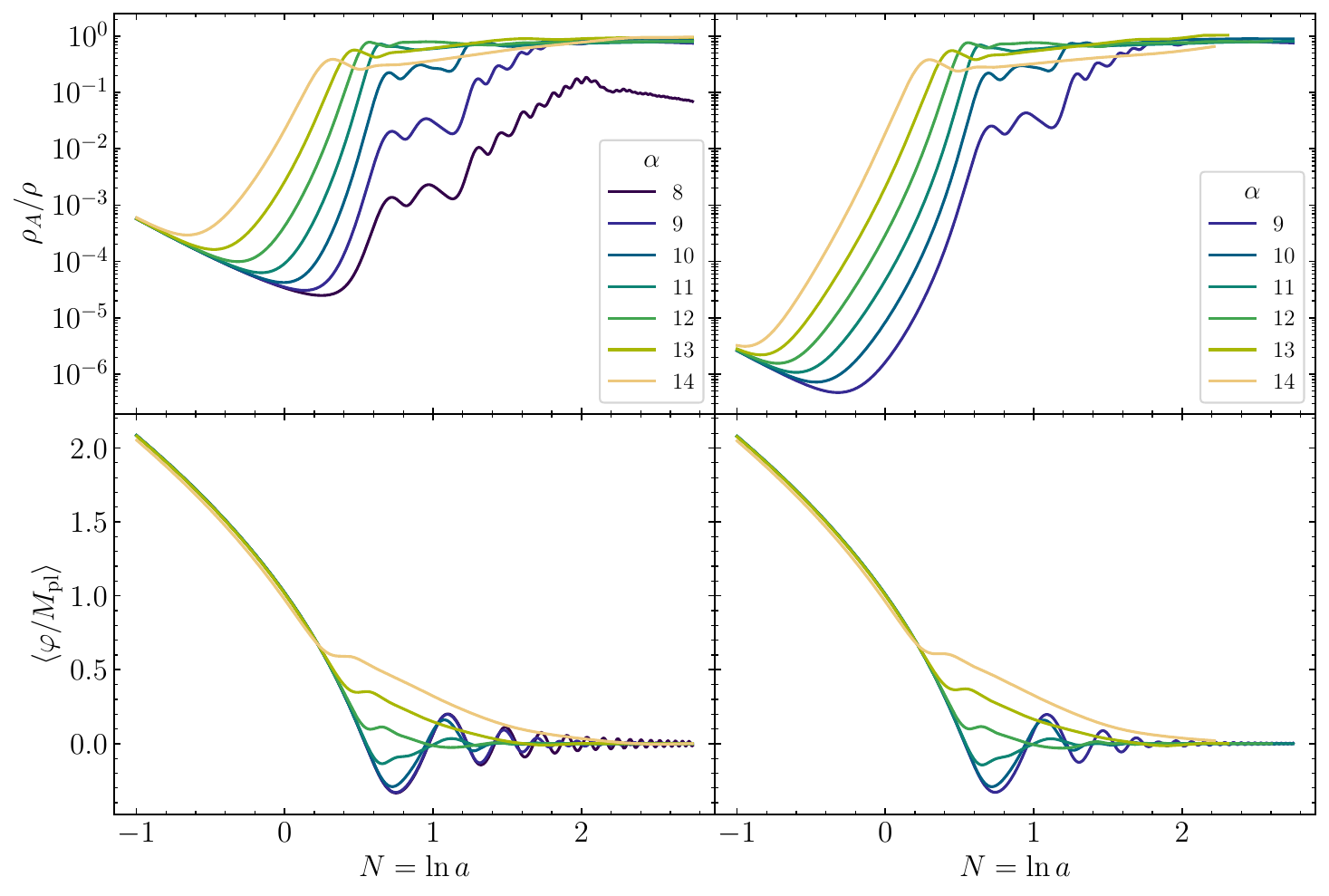}
    \caption{
        Energy fraction in the gauge fields (top panels) and homogeneous component of the inflaton (bottom panels)
        for FLRW simulations (left panels) and BSSN simulations (right panels).
        Note that at early times ($N \lesssim 0$) the gauge field energy densities differ between the two methods only due to the differing choice of cutoffs in initial conditions, as the dominant contribution is from vacuum modes (and unphysical).
    }
    \label{fig:background}
\end{figure}

The larger the axial coupling, the wider the instability---thereby increasing the efficiency by which energy is transferred from the homogeneous mode to the gauge fields.  At low couplings, $\alpha_g \lesssim 9$, the gauge fields never fully dominate the energy budget of the Universe and we consider preheating to be incomplete.  At couplings just above, with $9 \lesssim \alpha_g \lesssim 10$, the majority of the energy contained in the homogeneous mode of the inflation is transferred to the gauge fields, but the process takes several oscillations.  If $10 \lesssim \alpha_g \lesssim 12$, the homogeneous mode of the field breaks down during the first oscillation.  In this regime, there is a substantial amount of backreaction onto the modes of the inflaton, as well.  For the highest couplings, $\alpha_g \gtrsim 13$, resonance is so strong that the homogeneous mode of the field never crosses zero before it decays.  In these cases, backreaction onto the  inflaton field is suppressed. At the highest coupling, the backreaction stalls the evolution of the inflaton on its potential, and inflation briefly restarts.  These four regimes can be seen in \cref{fig:background} for the FLRW case (left panels) which show both the energy contained in the gauge field as a function of time and the mean of the inflation.  More details on this evolution can be found in \cite{Adshead:2019igv,Adshead:2019lbr}.
The right panel of \cref{fig:background} shows the same quantities when implemented in the BSSN scheme.  The inclusion of nonlinear gravity has little to no effect on the qualitative structure of preheating at the level of the background, for the wide range of parameters we have simulated.

\begin{figure}[ht]
    \centering
    \includegraphics[width=\textwidth]{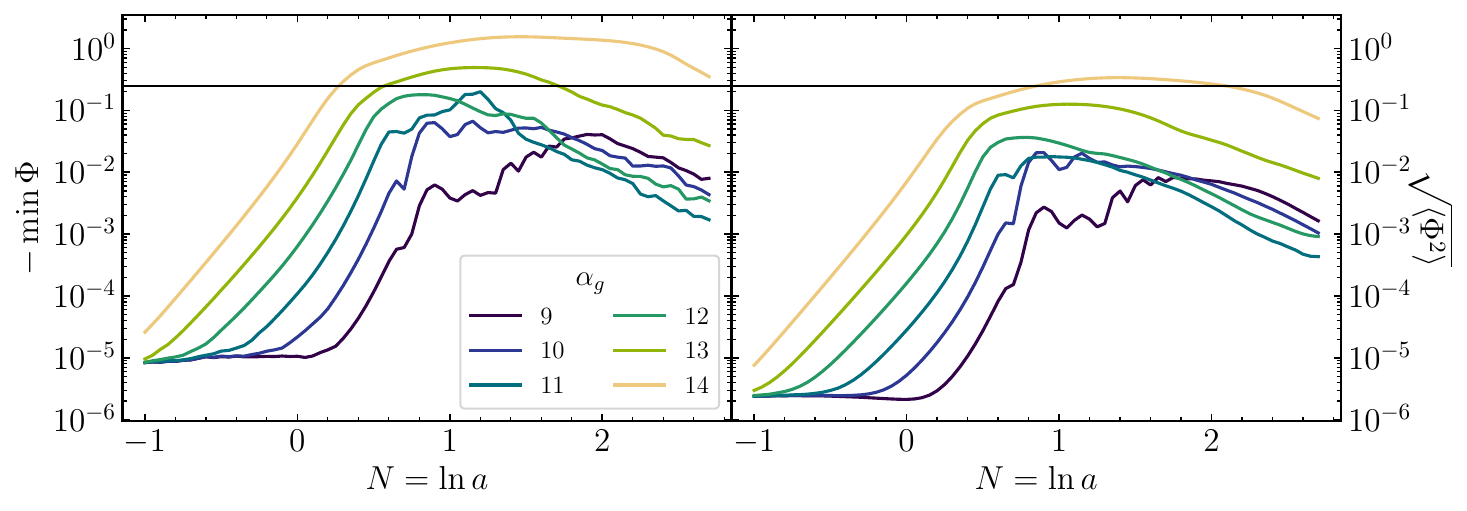}
    \caption{
        Statistics of the Newtonian potential calculated passively from FLRW simulations.  The left panel shows the absolute value of the minimum value of the Newtonian potential on each slice, the right panel shows the square root of the variance of the Newtonian potential across the grid.  The solid horizontal lines denote the value $\Phi = 0.25$, the point at which linearized gravity breaks down.
        Curves are not smooth only because these quantities are calculated at relatively infrequent
        intervals.
    }
    \label{fig:newtpot}
\end{figure}

Nonetheless, the system exhibits large density contrasts, especially for larger couplings.  As a first attempt, we can look for local gravitational effects using a linearized scheme.  In conformal Newtonian gauge, where the scalar part of the metric is
\begin{equation}
g^{\rm Newt}_{\mu\nu} = a(\tau)^2 \diag\left[-\left(1+2\Phi\right),\left(1-2\Phi\right),\left(1-2\Phi\right),\left(1-2\Phi\right)\right],
    \label{eq:newtgagemet}
\end{equation}
we can solve for the Newtonian potential $\Phi$ with the $00$ component and the (divergence of) the $0i$ components of the Einstein equations.
These respectively are
\begin{equation}
\label{poisson1}
    \partial_i \partial_i \Phi - 3 \mathcal{H} \left( \Phi^\prime + \mathcal{H} \Phi \right)
    = \frac{a^2}{2 \Mpl^2} \delta \rho
\end{equation}
and
\begin{equation}
\label{poisson2}
    \partial_i \partial_i \left( \Phi^\prime + \mathcal{H} \Phi \right)
    = - \frac{a^2}{2 \Mpl^2} \partial_i T_{0 i},
\end{equation}
which we solve using the same method as described in~\cite{Giblin:2019nuv}.  \Cref{fig:newtpot} shows the statistics of the Newtonian potential calculated from our FLRW simulations in \cref{fig:background}.  In this work, we calculate the Newtonian potential passively---with no feedback onto the evolution of the fields---to show that the Newtonian potential becomes too large, $\Phi >0.25$, to be able to treat gravity to linear order for the specific situations we consider.  At this value, $\sqrt{-g^{\rm Newt}} = a^4 \left(1-4\Phi\right)$ locally changes sign and linearized gravity, in conformal Newtonian gauge, necessarily breaks down.


\subsection{Density contrast and gravitational collapse}\label{subsec:density}

While the main features of the preheating story remain unchanged in the presence of nonlinear gravity, we now look to see if the additional nonlinear interactions provided by the gravitational sector affect the modes and scales that participate.  Specifically, we search for hints that these interactions may lead to gravitational collapse.

\begin{figure}[t!]
    \centering
    \includegraphics[width=\textwidth]{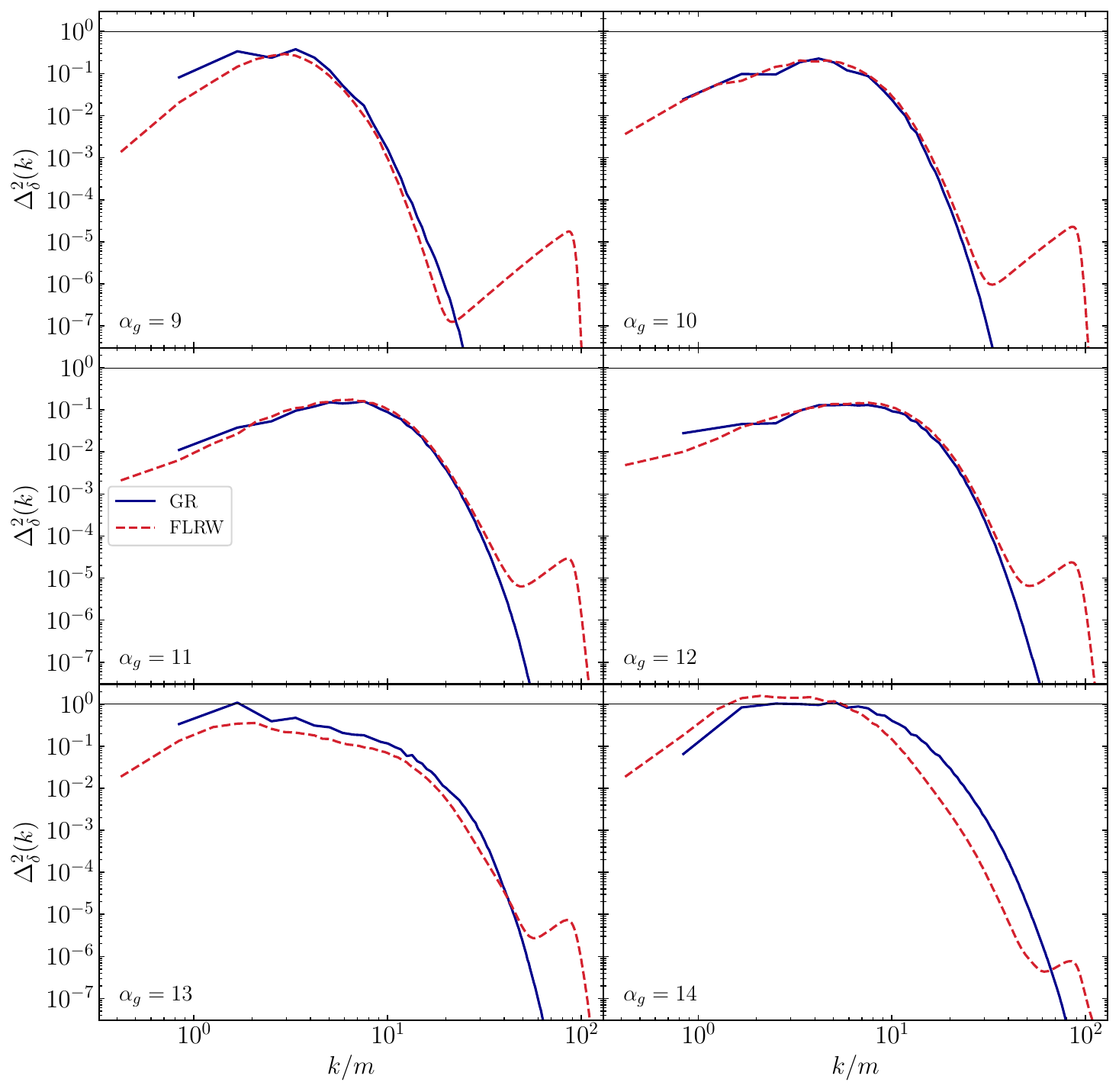}
    \caption{
        Spectra of density fluctuations $\delta = \delta \rho / \bar{\rho}$ in simulations
        with axial coupling $\alpha$ varying by panel, comparing results from simulations
        implementing full general relativity via the BSSN scheme (solid blue lines) and
        FLRW simulations (dashed red).
        All results are evaluated two $e$-folds after the end of inflation. The smaller, higher-frequency peak in the FLRW simulations are a numerical artifact that corresponds to the scale of the cutoff of the initial conditions.
    }
    \label{fig:delta-spectra-bssn-vs-flrw}
\end{figure}

The first place we look is at the power spectrum of density fluctuations.  In \cref{fig:delta-spectra-bssn-vs-flrw} we plot the dimensionless power spectrum of $\delta \equiv \delta \rho / \bar{\rho}$,
\begin{equation}
    \Delta^2_\delta = \frac{k^3}{2\pi^2}P_\delta(k),
\end{equation}
where the power spectrum is defined from the two point correlation function,
\begin{align}
    \left<\delta(k)\delta(k^\prime)\right> = \left(2\pi\right)^3\delta^3({\bf k}-{\bf k}^\prime) P_\delta(k).
\end{align}
When calculating $\Delta^2_\delta$ we directly Fourier Transform the $\rho$ as calculated in \cref{eqn:rho} on the hypersurfaces of the simulation and ignore spatial dependence of the conformal factor.  This measure is often cited as the litmus test for the need to incorporate nonlinear effects (see, for example, Ref.~\cite{Widrow:2009ru})  as well as a useful measure to determine whether primordial black holes are produced.  This statistic was first used as a way to diagnose potential primordial black hole formation in Ref.\ \cite{Carr:1975qj} and continues to be used, see, for example, Refs.~\cite{Martin:2019nuw,Ozsoy:2023ryl}.

\Cref{fig:delta-spectra-bssn-vs-flrw} shows that, in many cases, dimensionless power spectra are of order $\Delta_\delta^2(k) \sim 1$ for large $\alpha_g$.  This is reinforced by the fact that perturbation theory breaks down at these times; more precisely, in cases where $\delta(k)$ approaches unity, the linearized Einstein's equations are violated as we saw in \cref{fig:newtpot}~\cite{Giblin:2018ndw}.

As one would expect, the simulations with nonlinear gravity show more numerical effects at high wavenumber; nonetheless, the physical parts of the power spectra are very consistent with the FLRW counterparts.  The only exception to this is, perhaps, the very highest values of the self-coupling where there seems to be somewhat higher power at lower scales and somewhat lower power at intermediate scales. We speculate that this is due to the fact that local gravity can lead to some clumping in the nonlinear simulations.

\begin{figure}[t!]
    \centering
    \includegraphics[width=\textwidth]{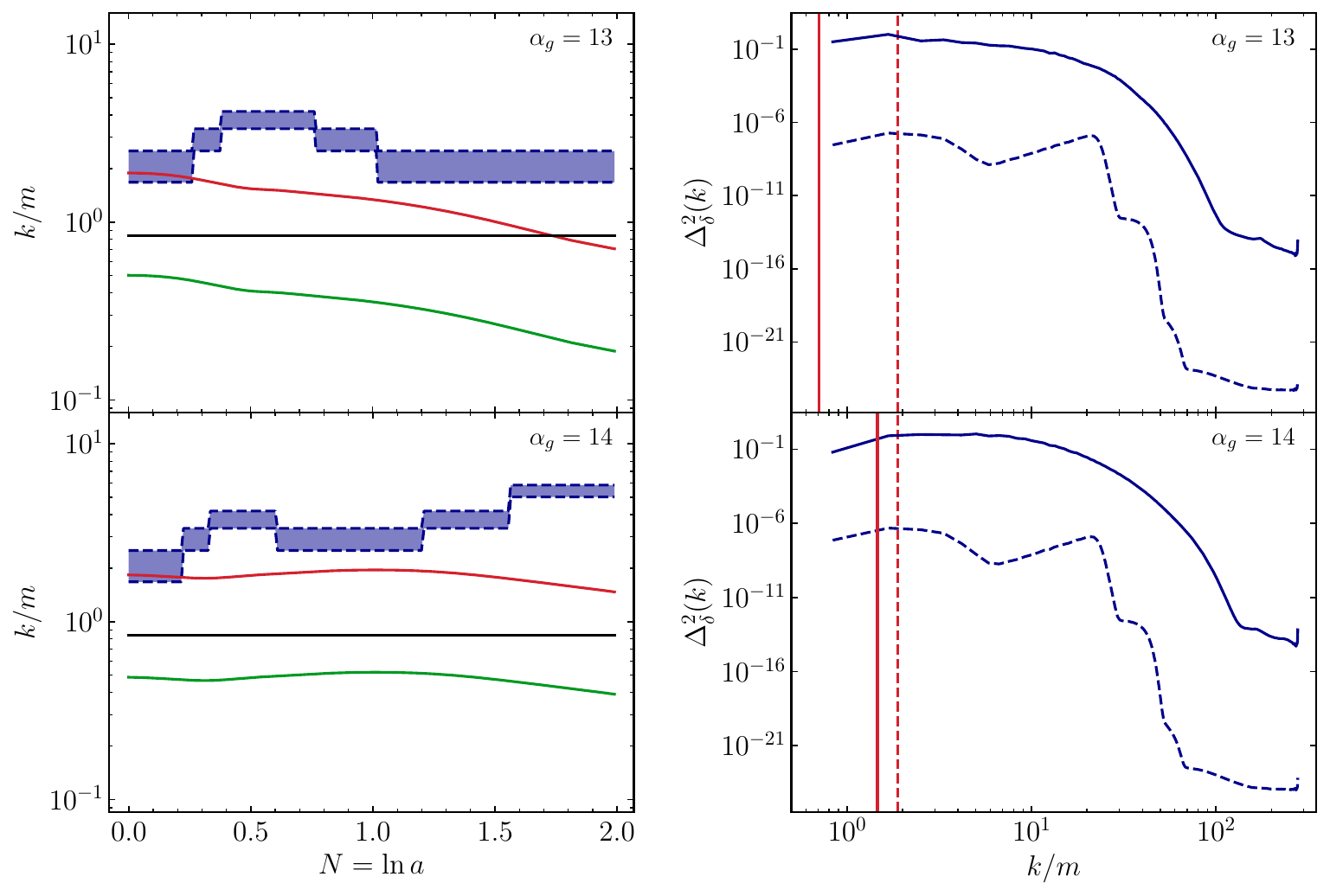}
    \caption{
        The comoving Jeans scale compared to the dimensionless power spectrum.  The left panels show the range of modes in the peak--most power--bin of the dimensionless power spectrum (blue, shaded region) compared to the Jeans scale (red), the Hubble scale (green), and the longest resolvable mode in the box (black).  We only calculate the location of the peak of the dimensionless power spectrum and the Jeans length after the end of inflation when the power spectrum raises above its initialized shape and we can approximate the Universe as radiation dominated.  The right panels show the dimensionless power spectrum at the end of inflation (blue, dashed lines) and two $e$-folds after the end of inflation (blue, solid lines) with vertical lines showing the Jeans corresponding jeans scale at the end of inflation (red, dashed lines) and two $e$-folds after the end of inflation (red, solid lines).  The top panels correspond to $\alpha_g = 13$ and the bottom panels correspond to $\alpha_g = 14$.
    }
    \label{fig:jeansplot}
\end{figure}

In order for clumps to actually collapse into black holes, overdensities must overcome their own internal pressure.  In linear theory, the {\sl Jeans} scale is used to determine whether the scales of interest can collapse. Modes with wavelength longer than the Jeans length  undergo gravitational collapse, while those with shorter wavelength are pressure supported. This scale defines to the (physical) Jeans scale,
\begin{equation}
    k_J^{\rm phys} = \frac{2\pi}{\lambda_J} = 2\pi\sqrt{\frac{\rho G}{c_s^2}} = \sqrt{\frac{3H^2}{2\pi c_s^2}},
\end{equation}
where $c_s^2 = \delta P/\delta \rho$ is the sound speed.  The comoving Jeans scale is $k_J = ak_J^{\rm phys}$.  Since the gauge fields are the main component of the universe, we estimate $c_s^2=1/3$ to approximate the sound speed for a radiation fluid.  At the end of our simulations---where the power spectra in fig.~\ref{fig:delta-spectra-bssn-vs-flrw} are evaluated---the Jeans scale $k_J$ is much smaller than the peak frequency of the dimensionless power spectrum of the density fluctuations. This indicates that the radiation pressure of the gauge fields is playing an important role in the evolution of these structures.  To use $\alpha_g = 13$ and $\alpha_g = 14$ as examples, we can track the location peak of the dimensionless power spectrum throughout simulation which we can compare to the Jeans scale and the Hubble scale as can be seen in fig.~\ref{fig:jeansplot}.  The differences in the evolution of the comoving
scales is due to the fact that, for the largest coupling, the gauge field backreaction briefly restarts inflation. Fig.~\ref{fig:jeansplot} shows us that Jeans-scale modes are excited near the end of inflation; however, these modes are still small.  As the density contrast grows, the peak of the power spectrum moves to slightly larger comoving wavenumber, while the comoving Jeans scale shrinks.  By the time we reach two $e$-folds after the end of inflation, the peak mode of the power spectrum  is approximately a factor of five larger than the Jeans scale.

\begin{figure}[t!]
    \centering
    \includegraphics[width=\textwidth]{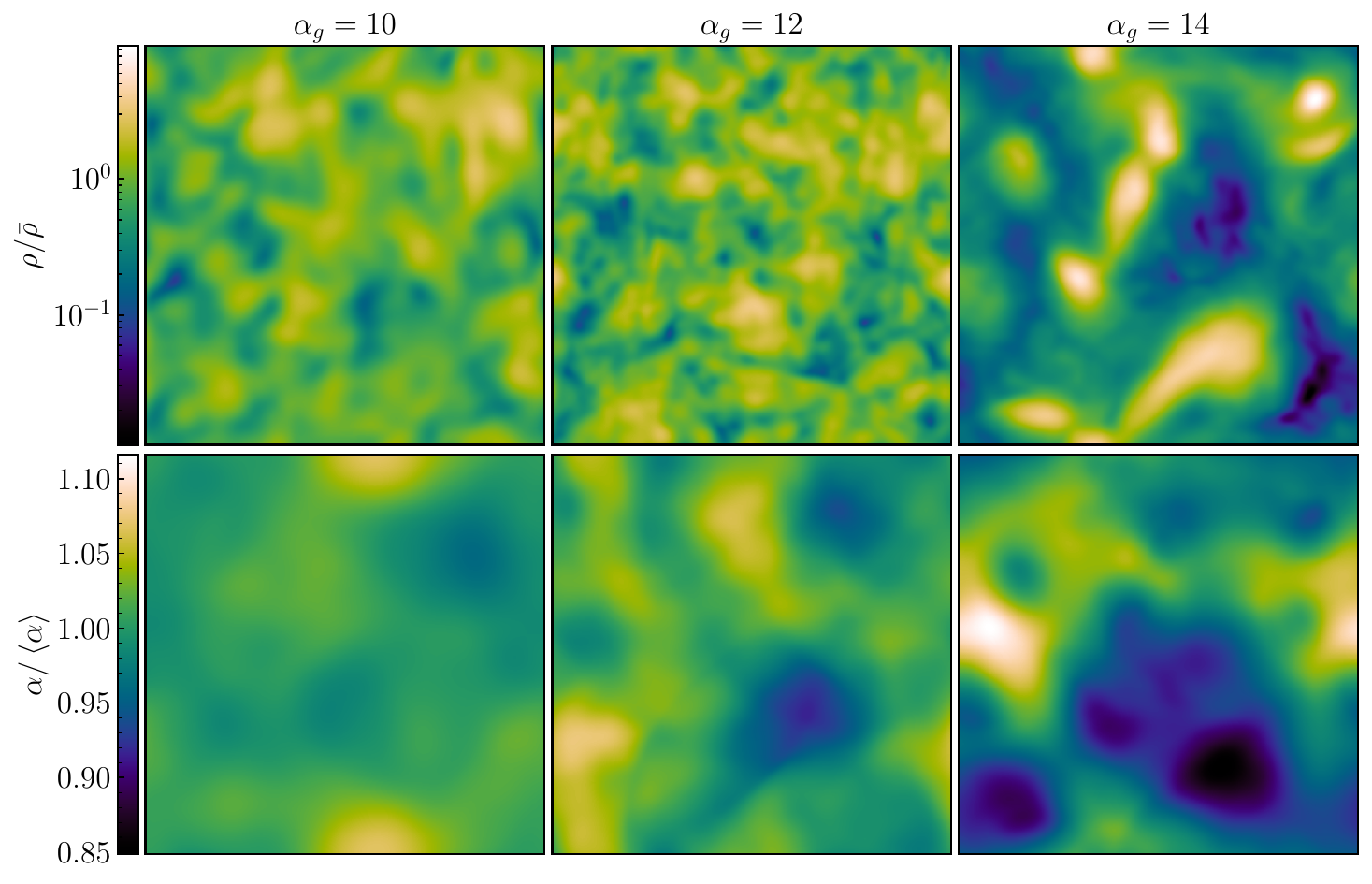}
    \caption{
        Two-dimensional spatial slices of the density contrast, $\delta$, (top panels) and
        lapse (normalized to its average over all space; bottom panels), $\alpha/\left<\alpha\right>$.
        Columns display results for simulations with axial coupling $\alpha_g = 10, 12$, and $14$
        from left to right.
        All results are evaluated two $e$-folds after the end of inflation.
    }
    \label{fig:delta-alpha-2d}
\end{figure}

\begin{figure}[t!]
    \centering
    \includegraphics[width=\textwidth]{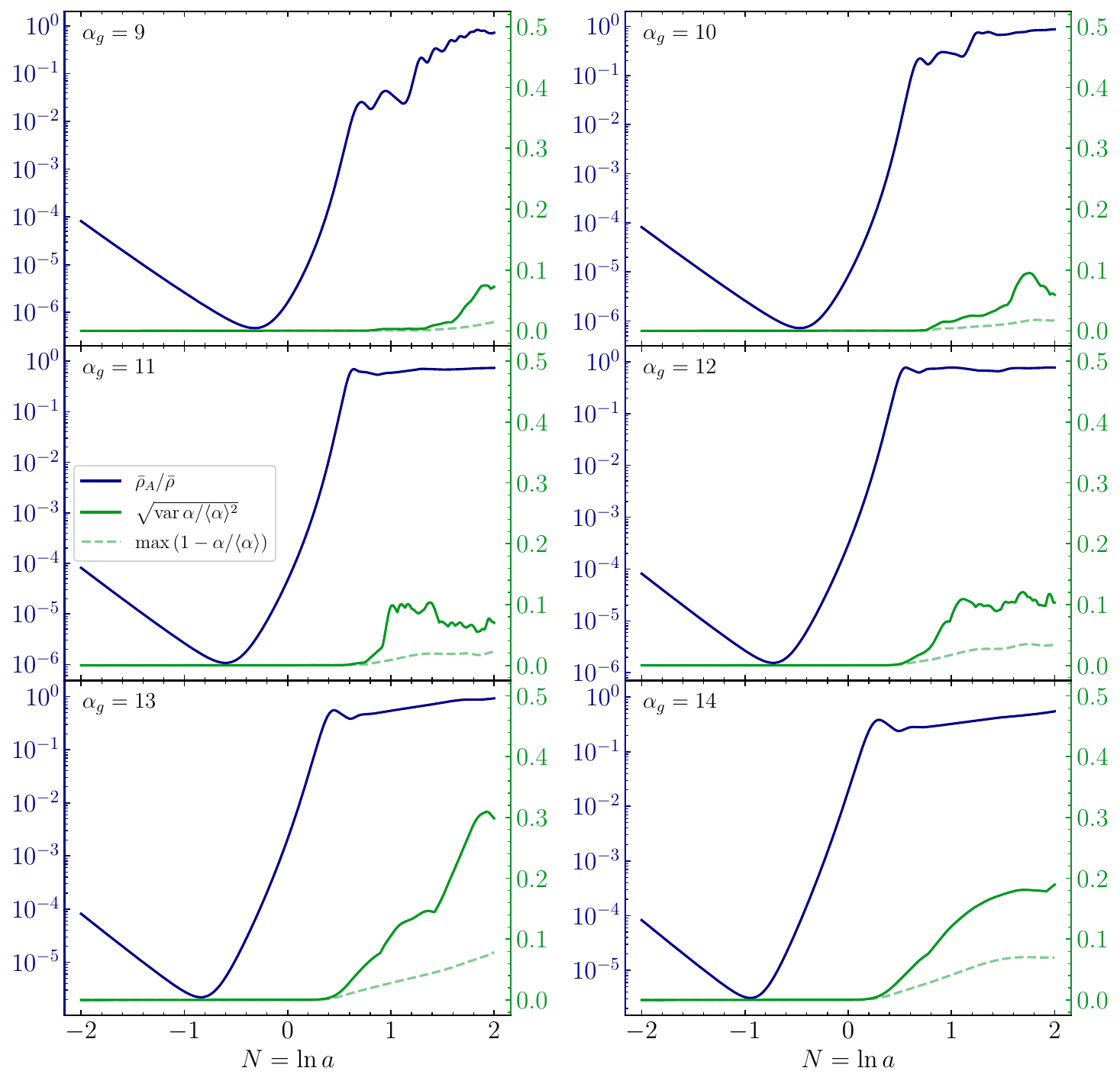}
    \caption{
        Dynamics of energy transfer and the metric lapse during preheating.
        Plotted is the fraction of energy in the gauge fields, $\bar{\rho}_A / \bar{\rho}$
        (solid blue, with scale given by the left of each panel) and the root-mean-squared
        and maximum deviation of the lapse from its average over space (solid green
        and transparent, dashed green, respectively, with scale given by the right of each panel).
        Each panel corresponds to a simulation with axial coupling strength indicated on the plot.
    }
    \label{fig:frac-gauge-alpha-stats}
\end{figure}

A number of different tools are used to identify the existence of black holes when studying critical collapse in numerical relativity~\cite{Baumgarte:2010ndz}.  In $1+\log$ slicing, where $(\partial_t + \beta^i\partial_i) \alpha = -2\alpha K$, the vanishing of the lapse $\alpha$ can sometimes be used as an indicator that black holes have formed~\cite{Hannam:2006vv,Hannam:2008sg,Bruegmann:2009hof,Hannam:2006xw,Baumgarte:2007ht,Dennison:2014sma,Baumgarte:2022auj}.  The same is true for our slicing condition, \cref{eq:lapse}, where the only difference is the background clock which is constructed to mimic a conformal-time FLRW solution in the homogeneous limit.  In the weak gravity limit, we can compute the inhomogeneity of $\alpha$ to measure how much our slices vary from the FLRW limit, with smaller values of $\alpha$ corresponding to deeper gravitational wells. \Cref{fig:delta-alpha-2d} shows 2-dimensional slices of three different simulations evaluated two e-foldings after the end of inflation. In these figures we can see spatial variation in the density contrast---which is often large, $\mathcal{O}(10)$.  The lapse, however, does not deviate much more than $\mathcal{O}\left(10^{-1}\right)$.
We can extend this to a statistical test by calculating the variance of the density contrast $\delta$ and the lapse $\alpha$ throughout the simulation.  As we can see in \cref{fig:frac-gauge-alpha-stats}, the variance of the lapse increases with the size of the axial coupling $\alpha_g$. However,  the deviation from the average is never larger than $\mathcal{O}(0.3)$.

As a final metric, we look to determine whether regions in our simulations have passed thresholds that one might consider sufficient to produce black holes.  We begin by examining just one of the more dramatic runs and calculating the compactness~\cite{Shibata:1999zs, Musco:2018rwt, Escriva:2019phb, Saini:2017tsz} generalized to an expanding spacetime \cite{Germani:2023ojx}
\begin{equation}
    C(R) = \frac{G\delta \! M}{R}
    \label{eq:compactness}
\end{equation}
to calculate whether the overdensities on the final slice indicate that black holes might form.  In \cref{eq:compactness} $\delta \! M$ is the instantaneous over-mass enclosed in some (proper) radius, $R$. That is, we subtract off the background density when computing the mass.
Note that we use a different definition of compactness than that in Ref.~\cite{Aurrekoetxea:2023jwd}, where the authors measure the total integrated $\rho$ within a (not necessarily spherical) region where $\rho/\bar{\rho} > 5\%$, although both definitions reduce to a statement of the {\sl hoop conjecture}~\cite{Misner:1973prb} about a spherically symmetric overdensity and when $\rho/\bar{\rho} \gg 1$.

\begin{figure}[t!]
    \centering
    \includegraphics[width=\textwidth]{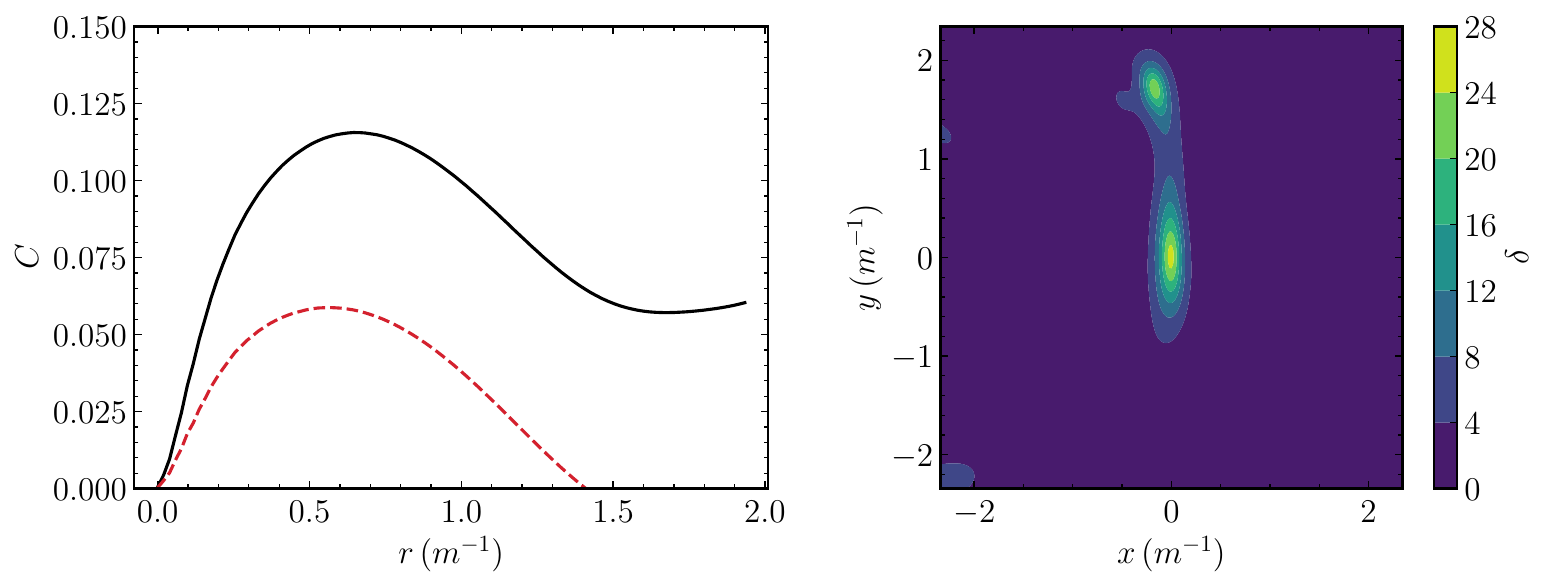}
    \caption{
       The left panel shows the compactness $C$ as a function of distance from the center of the largest overdensity present in the final slice of the simulation with $\alpha_g = 14$ (black, solid).  We also plot the FLRW approximation to the compactness (red, dashed).  The right panel shows the $z={\rm constant}$ slice through the center of the overdensity.
    }
    \label{fig:compactness}
\end{figure}

To test this, we look at the final slice for the largest coupling we test, $\alpha_g = 14$, and compute the compactness of the largest over-density in the box. Centering coordinates about the location of maximum $\delta(x)$, we can calculate the enclosed over-mass
\begin{equation}
\delta  \! M(R)  = \bar{\rho} \int_0^R \sqrt{\gamma} \left(\delta-1\right) \, r^2dr = \bar{\rho} \Delta x^3 \sum_i\sqrt{\gamma_i} \left(\delta(x_i)-1\right).
\label{eq:propmass}
\end{equation}
and, following \cite{Aurrekoetxea:2023jwd}, we approximate the radius of this overdensity from $V = 4R^3/3\pi$,
\begin{equation}
R = \left(\frac{3\pi}{4}\right)^{1/3}\left(\int_0^R \sqrt{\gamma} r^2 dr \right)^{1/3}=  \left(\frac{3\pi}{4}\right)^{1/3}\left(\Delta x^3 \sum_i\sqrt{\gamma_i} \right)^{1/3}.
\label{eq:propR}
\end{equation}
In both \cref{eq:propmass} and \cref{eq:propR} the final sum is over all points inside a given radius $r$.   \Cref{fig:compactness} shows the compactness as a function of distance away from the center---a measure that is maximized around $C\sim 10^{-1}$, yet is still smaller than the critical value of $1/2$. In \cref{fig:compactness} we also plot the FLRW approximation to the compactness where $ \delta \! M_{\rm FLRW} = a^3 \bar{\rho} \Delta x^3 \sum_i \left(\delta_i-1\right)$  and $R_{\rm FLRW} = a \Delta x$ where the scale factor is calculated in the FLRW limit, $a^2 = \bar{e^{4\phi}}$, see \cref{app:perttheory}.


\subsection{Gravitational waves}\label{subsec:GWs}

One of the most striking features of gauge preheating is the strength of the production of gravitational radiation. During and after gauge preheating, gravitational wave production can be so strong that the resulting radiation density stored in gravitational waves leads to a shift in the effective number of relativistic species $N_{\rm eff}$ that is large enough to be measured or constrained by current and future CMB experiments~\cite{Adshead:2019lbr,Adshead:2019igv,Adshead:2018doq}. In some models, the shift is large enough that the allowed values of the axial coupling are already constrained by existing CMB measurements. These results were obtained using numerical simulations that ignore the backreaction of these large metric fluctuations. In this section, we explore the robustness of predictions of gravitational wave production during gauge preheating in the regime where gravity is nonlinear.

Gravitational waves are the traceless part of the spatial metric, and are contained in $\bar{\gamma}_{ij}$. In FLRW simulations, the gravitational wave spectrum is computed passively---see, e.g.~\cite{Easther:2007vj}, by calculating the transverse-traceless parts of the matter stress tensor,
\begin{equation}
    T_{ij}^{\rm TT} = \left(P_{il} P_{jm} - \frac{1}{2} P_{ij} P_{lm} \right) T_{lm},
\end{equation}
where $P$ is the projection operator,
\begin{equation}
	P_{ij} = \delta_{ij} - \frac{k_i k_j}{k^2}.
\end{equation}
This sources the transverse-traceless part of the metric via Einstein's equations
\begin{align}
\Box h_{ij} = 16 \pi G \,T^{\rm TT}_{ij}.
\end{align}
We can then calculate the power in gravitational waves from the gravitational wave stress-energy tensor~\cite{Misner:1973prb}
\begin{equation}
    T_{\mu \nu}^{\rm GW} = 8\pi G \left<h_{ij,\mu}^{\rm TT}{{h^{ij}}_{\nu}}^{\rm TT}\right>
\end{equation}
where
\begin{align}
\rho_{\rm gw}^{\rm FLRW} = 8\pi G \left|h_{ij,0}^{\rm TT}\right|^2.
\end{align}
The energy density in gravitational waves at the time of emission is then given by
\begin{equation}
\label{omegagw}
    \Omega^{\rm FLRW}_{\rm gw}(k)
    \equiv \frac{1}{\rho} \frac{\ud \rho_\mathrm{gw}}{\ud \ln k}
	= \frac{1}{24\pi^2 L^3} \frac{k^3}{\mathcal{H}^2} \sum_{i, j} \left\vert h_{ij}^\prime(k, \tau)) \right\vert^2.
\end{equation}

Since the BSSN system evolves the entire metric by solving the full nonlinear Einstein equations, the gravitational wave spectrum can be directly extracted by projecting out the transverse-traceless part of the extrinsic curvature, $h_{ij}^\prime \approx -2\tilde{A}_{ij}$~\cite{Bastero-Gil:2010tpb}.  The BSSN analog to eq.\ \eqref{omegagw} is then
\begin{align}
    \Omega^{\rm BSSN}_{\rm gw}(k)
    &= \frac{3}{2 \pi^2 L^3} \frac{k^3}{\left\langle K \right\rangle^2}
        \sum_{i, j} \left\vert \tilde{A}_{ij}(k, t) \right\vert^2,
\end{align}
which is analogous to the quantities used in Refs.~\cite{Kou:2021bij,Ota:2021fdv}.

The spectral energy density of gravitational waves today can be calculated using the standard transfer functions~\cite{Easther:2006gt,Easther:2006vd}, assuming that the signal is evaluated during a radiation-dominated period and remains radiation dominated until matter-radiation equality.  The spectral energy density today is given by
\begin{align}
    \Omega_{\mathrm{GW}, 0} h^2
    &= \Omega_{\mathrm{rad}, 0} h^2
        \left( \frac{g_{\star S}(a_0)}{g_{\star S}(a_\mathrm{r})} \right)^{1/3}
        \Omega_\mathrm{GW}(a),
\end{align}
and the frequency by
\begin{align}
    f
    &= \frac{k / 2 \pi a}{\rho(a)^{1/4}}
        \rho_{\mathrm{r}}(a_0)^{1/4}
        \left( \frac{g_{\star}(a_\mathrm{r})}{g_{\star}(a_0)} \right)^{1/4}
        \left( \frac{g_{\star S}(a_\mathrm{r})}{g_{\star S}(a_0)} \right)^{-1/3}
    \\
    &= 3.2 \times 10^{10} \, \mathrm{Hz} \frac{k / a}{\sqrt{H(a) \Mpl}}
        \left( \frac{g_{\star}(a_\mathrm{r}) / g_{\star}(a_0) }{100} \right)^{-1/12}.
\end{align}
In the preceding expressions $g_\star$ is the number of ultra-relativistic degrees of freedom evaluated at reheating ($a_r$) or today ($a_0$).  For simplicity, we take $g_{\star}(a_\mathrm{r})/g_{\star}(a_0)\approx 100$.

\begin{figure}[t!]
    \centering
    \includegraphics[width=\textwidth]{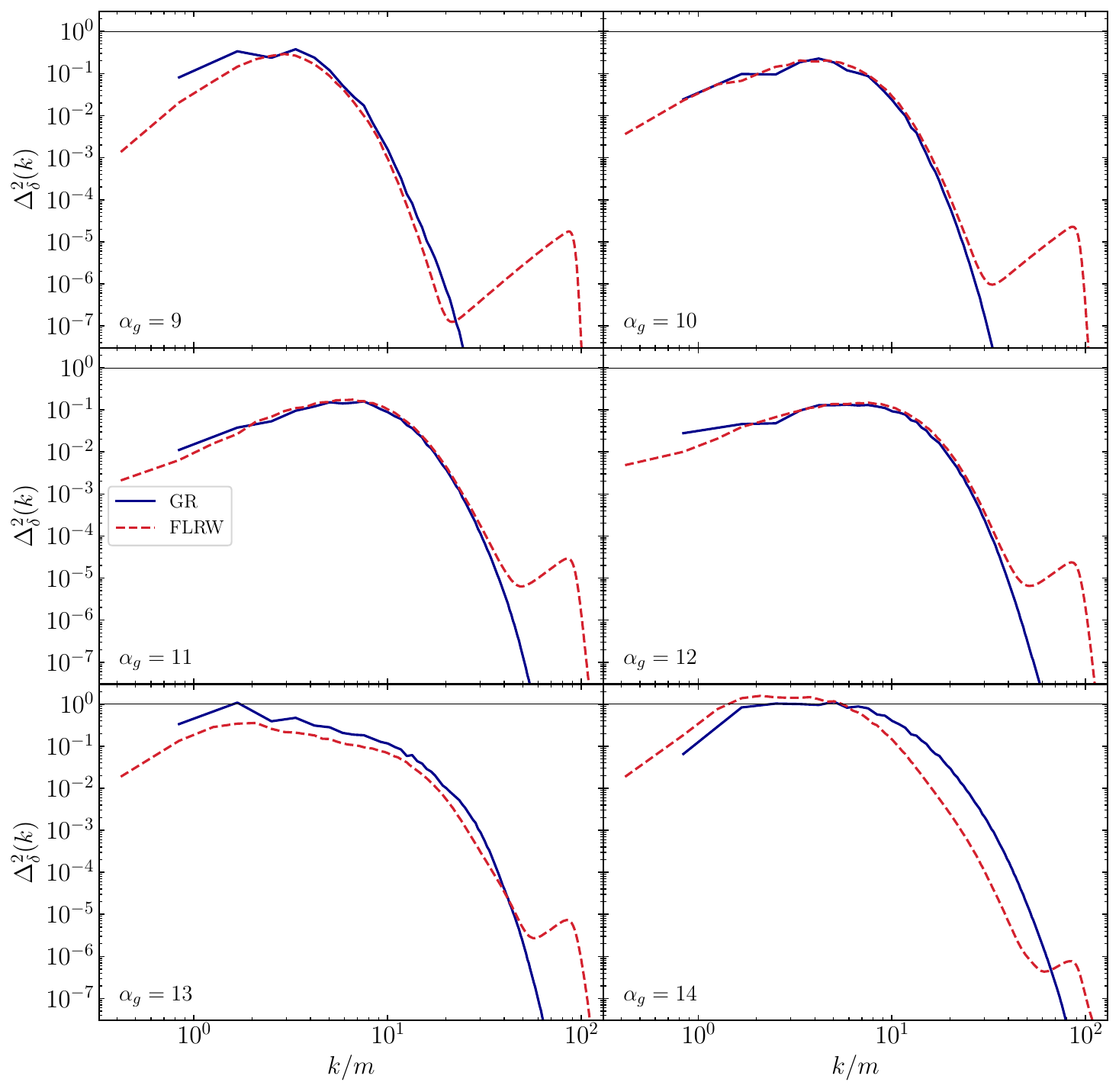}
    \caption{
        Present-day gravitational wave spectra in simulations
        with axial coupling $\alpha_g$ varying by panel, comparing results from simulations
        implementing full general relativity via the BSSN scheme (solid blue lines) and
        FLRW simulations (dashed red).
        All results are evaluated two $e$-folds after the end of inflation.
    }
    \label{fig:gw-bssn-vs-flrw}
\end{figure}

In \cref{fig:gw-bssn-vs-flrw} we show the gravitational wave spectra from simulations of gauge preheating in both full nonlinear gravity, as well as in rigid FLRW spacetime. Despite the presence of strong nonlinearities, we observe remarkably consistent outputs across the range of couplings we considered. These results suggest that neglecting nonlinear gravity has at most an $\mathcal{O}(1)$ effect on the resulting gravitational wave spectrum, even in regions where the metric is highly nonlinear.


\section{Conclusions}
\label{sec:conclusions}

In this paper we have extended studies of gauge preheating after pseudoscalar-driven inflation to include the effects of nonlinear gravitation. To facilitate this study, we implemented numerical relativity using the BSSN formalism in our simulation software.

The evolution of the energy density and fields was nearly indistinguishable from simulations done in a FLRW spacetime.  Including the effects of nonlinear gravity made no qualitative difference in the value or evolution of the averaged background quantities, such as the expansion rate, the average value of the scalar field, or the energy densities of the scalar and gauge fields.

In our FLRW simulations we observed the emergence of regions with very large fractional overdensities $\delta\rho/\rho$, which routinely exceeded unity. The existence of these regions leads to the failure of linearized gravity, requiring software that properly treats evolution of the metric into the nonlinear regime. In our BSSN simulations, we observe power being shifted from large to small scales due to the gravitational interactions.

In our highest-coupling BSSN runs, we found regions with fractional overdensities as large as $\delta\rho/\rho \sim 30$. However, despite the development of these large density contrasts we found no evidence for the formation of black holes such as the presence of horizons or the vanishing of the lapse function. Closer examination of these simulations revealed that the scales where the density power spectrum peaks are within the Jeans length. This suggests that pressure of the radiation plays an important role in the evolution and stability or instability of these very overdense regions. We also computed the compactness of these regions, and found that it peaks at values $C\sim 10^{-1}$, which is somewhat below the level required to make a black hole of $C\sim 0.5$.

We also studied the resulting spectrum of gravitational waves in these scenarios. A key result of our earlier work was the prediction of a very large gravitational wave signal at the strongest couplings. Our simulations in full nonlinear gravity revealed the robustness of our original FLRW simulations of gravitational wave production.

Looking further ahead, our BSSN simulations of gauge preheating end sooner than would be ideal, owing to the large increase in the volume of the box over the course of four $e$-foldings of expansion and the movement of power to large wave number after this time. We plan to investigate the dynamics of the large overdense regions to study their fate. In particular, it would be interesting to investigate whether the large overdensities subsequently undergo gravitational collapse, or whether they simply decay. Our simulations have so far focused on the large density fluctuations that are generated on sub-horizon scales during reheating. These are necessarily inside the Jeans length at production. It would be interesting to study the horizon reentry of large curvature perturbations that are produced near the end of inflation in these models to see if their collapse to black holes can be verified in full general relativity.

\acknowledgments

We thank Thomas Baumgarte, Katy Clough, Valerio de Luca, Daniel Figueroa, Carsten Gundlach, Eugene Lim and Andrew Tolley for useful discussions.

P.A.\ is supported by the United States Department of Energy, DE-SC0015655.
J.T.G.\ is supported by the National Science Foundation, PHY-2309919.
Z.J.W.\ is supported by the Department of Physics and the College of Arts and Sciences at the
University of Washington. P.A. thanks the Center for Particle Cosmology at the University of Pennsylvania for hospitality while this work was being completed. P.A., J.T.G., and Z.J.W. gratefully acknowledge support from the Simons Center for Geometry and Physics, Stony Brook University at which some of the research for this paper was performed.
We acknowledge the  National Science Foundation, Kenyon College and the Kenyon College Department of Physics for providing the hardware used to carry out these
simulations.
This work used the Extreme Science and Engineering Discovery Environment (XSEDE)~\cite{xsede}, which
is supported by National Science Foundation grant number ACI-1548562; simulations were run through
allocation TG-PHY200037 on Bridges-2 at the Pittsburgh Supercomputing Center which is supported by
NSF award number ACI-1928147, and Expanse at the San Diego Supercomputer Center.

The FLRW simulations presented here are implemented with \textsf{pystella}~\cite{pystella}, which
is available at \href{https://github.com/zachjweiner/pystella}{github.com/zachjweiner/pystella} and
makes use of the Python packages \textsf{PyOpenCL}~\cite{kloeckner_pycuda_2012},
\textsf{Loopy}~\cite{kloeckner_loopy_2014},
\textsf{mpi4py}~\cite{DALCIN2008655,DALCIN20051108,Dalcin_mpi4py_Status_Update_2021},
\textsf{mpi4py-fft}~\cite{jpdc_fft}, and \textsf{NumPy}~\cite{Harris:2020xlr}.
This work also made use of the packages \textsf{SciPy}~\cite{Virtanen:2019joe},
\textsf{matplotlib}~\cite{Hunter:2007ouj}, \textsf{SymPy}~\cite{Meurer:2017yhf}, and
\textsf{CMasher}~\cite{cmasher}.


\appendix


\section{BSSN decomposition}


\label{app:BSSNdeets}

In this appendix we present the details of our decomposition of the gauge fields in the BSSN formalism.  We begin by detailing the $3+1$ decomposition of the metric and the foliation of the spacetime in \cref{3p1decomp}, before detailing the equations of motion for the scalar and gauge fields in \cref{appsec:EOMS}, and the derivation of the stress-energy tensor and sources for the evolution of the BSSN system in \cref{appsec:set}.

\subsection{3 + 1 decomposition}
\label{3p1decomp}
The line element in $3+1$ form is
\begin{align}
    \ud s^2
    &= - \alpha^2 \ud t^2
        + \gamma_{ij}
            \left( \ud x^i + \beta^i \ud t \right)
            \left( \ud x^j + \beta^j \ud t \right),
\end{align}
where the lapse $\alpha$ and shift $\beta^i$ parameterize gauge degrees of freedom, while the spatial metric $\gamma_{ij}$
and $\gamma^{ij}$ are used to lower and raise indices of spatial tensors.
The lapse and shift define the normal vector to hypersurfaces,
\begin{align}\label{eqn:normal-vector-def-app}
    n^\mu = \frac{1}{\alpha} \left(1, - \beta^i \right).
\end{align}
From the normalization condition of the normal vector, $n_\mu n^\mu = -1$, we can calculate its  inverse,
\begin{align}\label{eqn:normal-vector-inverse}
    n_\mu
    &= (- \alpha, 0, 0, 0).
\end{align}
The projector onto spatial hypersurfaces is
\begin{align}\label{eqn:projection-tensor-def}
    \gamma^\mu_{\hphantom{\mu}\nu}
    = \delta^\mu_{\hphantom{\mu}\nu} + n^\mu n_\nu.
\end{align}

The 3-dimensional covariant derivative is
\begin{align}
    D_\mu f
    \equiv \gamma_\mu^{\hphantom{\mu}\nu} \nabla_\nu f,
\end{align}
which is expressed via the 3-dimensional connection coefficients,
\begin{align}
    \Gamma^i_{\hphantom{i}jk}
    = \frac{1}{2} \gamma^{il} \left( \partial_k \gamma_{lj} + \partial_j \gamma_{lk} - \partial_l \gamma_{jk} \right).
\end{align}
The extrinsic curvature tensor is
\begin{align}\label{eqn:def-extrinsic-curvature}
    K_{\mu \nu}
    &= - \gamma_\mu^{\hphantom{\mu}\alpha} \gamma_\nu^{\hphantom{\nu}\beta} \nabla_\alpha n_\beta,
\end{align}
and its trace is
\begin{align}\label{eqn:trace-extrinsic-curvature-div-n}
    K
    \equiv g^{\mu \nu} K_{\mu \nu}
    &= - \nabla^\alpha n_\alpha.
\end{align}


\subsection{Equations of motion}\label{appsec:EOMS}

For completeness, we generalize the action \cref{eqn:lagrangian} to include a kinetic coupling $W(\varphi)$,
\begin{align}\label{eqn:lagrangian-appendix}
    S &= \int \ud^4 x \, \sqrt{-g} \left[
        \frac{\Mpl^2}{2} R- \frac{1}{2} \nabla_\mu \varphi \nabla^\mu \varphi
        - V(\varphi)
        - \frac{W(\phi)}{4} F_{\mu\nu} F^{\mu\nu}
        - \frac{X(\varphi)}{4} F_{\mu\nu} \tilde{F}^{\mu\nu}\right].
\end{align}

\subsubsection{Scalar fields}

Starting with the action \cref{eqn:lagrangian-appendix}, the Euler-Lagrange equation for the scalar field is
\begin{align}\label{eqn:scalar-ele}
    - \nabla_\mu \nabla^\mu \varphi
    = \pd{\mathcal{L}}{\varphi}
    &= - \dd{V}{\varphi}
        - \frac{1}{4} \dd{W}{\varphi} F_{\mu \nu} F^{\mu \nu}
        - \frac{1}{4} \dd{X}{\varphi} F_{\mu \nu} \tilde{F}^{\mu \nu}.
\end{align}
To rewrite this in the $3+1$ form, we must first decompose $\nabla_\mu \varphi$ into its parts normal to ($\Pi$) and lying in ($D_\mu \varphi$) spatial hypersurfaces,
\begin{align}
    \nabla_\nu \varphi
    &= D_\nu \varphi - n_\nu \Pi.
\end{align}
This defines the scalar's conjugate momentum,
\begin{align}\label{eqn:Pi-def-app}
    \Pi
    &\equiv n^\mu \nabla_\mu \varphi
    = \frac{1}{\alpha} \left(
        \nabla_0 \varphi - \beta^k \nabla_k \varphi
    \right).
\end{align}
A similar expansion of the covariant d'Alembertian yields
\begin{align}
    \nabla^\mu \nabla_\mu \varphi
    &= \frac{1}{\alpha} \gamma^{ij} D_i \alpha D_j \varphi
        + \gamma^{ij} \left(
            \partial_i D_j \varphi
            - \Gamma^{k}_{\hphantom{k}ij} D_k \varphi
        \right)
        + K \Pi
        - \frac{1}{\alpha} \left(\partial_t \Pi - \beta^k D_k \Pi \right).
\end{align}
Rearranging \cref{eqn:scalar-ele} into an explicit evolution equation for $\Pi$,
\begin{align}\label{eqn:final-Pi-eom}
    \partial_t \Pi
    &= \beta^k D_k \Pi
        + \gamma^{ij} \left(
            \alpha \partial_i D_j \varphi
            + D_i \alpha D_j \varphi
        \right)
        + \alpha \left(
            K \Pi
            - \gamma^{ij} \Gamma^k_{\hphantom{k}ij} D_k \varphi
            - \pd{\mathcal{L}}{\varphi}
        \right),
\end{align}
while solving \cref{eqn:Pi-def-app} for $\partial_t \varphi$ yields
\begin{align}\label{eqn:varphi-eom}
    \partial_t \varphi = \beta^k D_k \varphi + \alpha \Pi.
\end{align}

To increase numerical stability, we promote the spatial derivatives of the scalar field to dynamical degrees of freedom
themselves,
\begin{align}
    \psi_i \equiv D_i \varphi.
\end{align}
These evolve according to the gradient of \cref{eqn:varphi-eom},
\begin{align}
    \partial_t \psi_i
    &= \beta^k \partial_k \psi_i
        + \psi_k \partial_i \beta^k
        + \alpha D_i \Pi
        + \Pi D_i \alpha.
\end{align}
Using this variable, \cref{eqn:final-Pi-eom} becomes
\begin{align}\label{eqn:final-Pi-eom-subs-psi}
    \partial_t \Pi
    &= \beta^k D_k \Pi
        + \gamma^{ij} \left(
            \alpha \partial_i \psi_j
            + D_i \alpha \psi_j
        \right)
        + \alpha \left(
            K \Pi
            - \gamma^{i j} \Gamma^k_{\hphantom{k} i j} \psi_k
            - \pd{\mathcal{L}}{\varphi}
        \right),
\end{align}
which gives us the full system, \cref{eq:eomsphi}.

\subsubsection{Gauge fields}\label{appsec:GFEOMS}

Again, starting with the model \cref{eqn:lagrangian-appendix}, the Euler-Lagrange equations for the gauge field are
\begin{align}
    \label{eqn:maxwell-equation-bssn}
	0 &= W(\varphi) \nabla^\alpha F_{\alpha\beta} + \partial^\alpha W(\varphi) F_{\alpha\beta} + \partial^\alpha X(\varphi) \tilde{F}_{\alpha\beta},
\end{align}
which may be rearranged into canonical form as
\begin{align}\label{eqn:maxwell-equation-canonical-form}
	\nabla_\mu F^{\mu \nu}
    &= - \frac{1}{W(\varphi)} \left(
            \partial_\mu W(\varphi) F^{\mu \nu}
            + \partial_\mu X(\varphi) \tilde{F}^{\mu \nu}
        \right)
    \equiv - J^\nu.
\end{align}
In the last line we define the ``source vector'' $J^\nu$ to encode the coupling to the
scalar.

To recast these into the BSSN formalism, we begin by splitting the vector potential $A_\mu$ into its components along and orthogonal to spatial
hypersurfaces,
\begin{align}\label{eqn:A-mu-decomposition}
    A_\mu
    \equiv \mathcal{A}_\mu + n_\mu \mathcal{A},
\end{align}
where $\mathcal{A}_\mu = \gamma_\mu^{\hphantom{\mu}\nu} A_\nu$ and $\mathcal{A} = - n^\nu A_\nu$.
The electric and magnetic fields are
\begin{align}
    E^\mu
    &= \gamma^\mu_{\hphantom{\mu} \nu} n_\alpha F^{\nu \alpha}
    \label{eqn:def-E-mu-app}
    \\
    B^\mu
    &= - \gamma^\mu_{\hphantom{\mu} \nu} n_\alpha \tilde{F}^{\nu \alpha}
    \label{eqn:def-B-mu-app} \\
    &= \frac{1}{2} \epsilon^{\mu \alpha \rho \gamma} n_\alpha F_{\gamma \rho}.
    \label{eqn:def-B-mu-ito-epsilon}
\end{align}
In terms of these variables, the field tensor is
\begin{align}
    F_{\mu \nu}
    &= n_\mu E_\nu - n_\nu E_\mu
        + D_\mu \mathcal{A}_\nu - D_\nu \mathcal{A}_\mu,
    \label{eqn:field-tensor-in-terms-of-E-and-A}
\end{align}
or equivalently
\begin{align}
    F^{\mu \nu}
    &= n^\mu E^\nu - n^\nu E^\mu
        + n_\sigma \epsilon^{\sigma \mu \nu \rho} B_\rho.
    \label{eqn:field-tensor-in-terms-of-E-and-B}
\end{align}
We also decompose the source vector analogously to the gauge field, \cref{eqn:A-mu-decomposition}:
\begin{align}
    J^\mu
    &\equiv \mathcal{J}^\mu + n^\mu \mathcal{J}.
\end{align}

We can extend the gauge field system to include a constraint damping field $Z$
as~\cite{Palenzuela:2009hx, Hilditch:2013sba, Zilhao:2015tya, Clough:2022ygm}
\begin{subequations}
\label{eqn:extended-maxwell}
\begin{equation}
     \nabla_\mu F^{\mu \nu} + J^\nu
        - \frac{1}{W(\varphi)} \left( \nabla^\nu Z - \kappa n^\nu Z \right) =0
    \label{eqn:extended-maxwell-eqn}
\end{equation}
and
\begin{equation}
     \nabla^\mu A_\mu
        + Z = 0.
    \label{eqn:extended-lorenz-gauge}
\end{equation}
\end{subequations}
We can now set Lorenz gauge, $\nabla^\mu A_\mu + Z = 0$, to obtain the evolution equation for
$\mathcal{A}$.
Expanding the covariant divergence in terms of the $3+1$ decomposition and rearranging gives
\begin{align}
    \partial_t \mathcal{A}
    &= \beta^k D_k \mathcal{A}
        + \alpha \left( K \mathcal{A} - D^i \mathcal{A}_i - Z \right)
        - \mathcal{A}_i D^i \alpha.
\end{align}
The dynamics of $\mathcal{A}_i$ follow from the definition of $E_i$:
\begin{align}
    \partial_t \mathcal{A}_i
    &= \beta^k \partial_k \mathcal{A}_i
        + \mathcal{A}_k \partial_i \beta^k
        - \mathcal{A} D_i \alpha
        - \alpha \left( E_i + D_i \mathcal{A} \right).
\end{align}
The electric field's own evolution equation comes from the projection of Maxwell's equation
\cref{eqn:extended-maxwell-eqn} onto spatial hypersurfaces,
\begin{align}
    \partial_t E^i
    &= \beta^k \partial_k E^i
        - E^k \partial_k \beta^i
        + \alpha \left( K E^i - \mathcal{J}^i + \frac{1}{W(\varphi)} D^i Z \right)
        + \epsilon^{i j k} D_j \left( \alpha B_k \right).
\end{align}
Note that the latter term may be rewritten as
\begin{align}
    \epsilon^{i j k} D_j \left( \alpha B_k \right)
    &= D_j \left(
            \alpha
            D_i \mathcal{A}_j
            - D_j \mathcal{A}_i
        \right).
\end{align}
The spatial component of the source vector evaluates to
\begin{align}
    \mathcal{J}^i
    &= \frac{1}{W(\varphi)} \left(
            \dd{W}{\varphi}
            \left[
                \Pi E^i
                - \epsilon^{i j k} D_j \varphi B_k
            \right]
            + \dd{X}{\varphi}
            \left[
                - \Pi B^i
                - \epsilon^{i j k} D_j \varphi E_k
            \right]
        \right).
\end{align}

Finally, when the constraint damping field $Z = 0$, the component of
\cref{eqn:maxwell-equation-bssn} normal to spatial hypersurfaces yields Gauss's law,
\begin{align}\label{eqn:gauss-law-no-Z}
    D_i E^i
    &= \mathcal{J},
\end{align}
where
\begin{align}
    \mathcal{J}
    &= - \frac{1}{W(\varphi)} \left(
            \dd{W}{\varphi} E^i D_i \varphi
            - \dd{X}{\varphi} B^i D_i \varphi
        \right).
\end{align}
When including the constraint damping field, this equation instead specifies the dynamics of $Z$.
Namely, \cref{eqn:gauss-law-no-Z} reads
\begin{equation}
    D_\mu E^\mu
        - \mathcal{J}
        - \frac{1}{W(\varphi)} \left(
            \frac{1}{\alpha}
            \left[
                \partial_t Z - \beta^k D_k Z
            \right]
            + \kappa Z
        \right) = 0,
\end{equation}
yielding the evolution equation
\begin{align}
    \partial_t Z
    &= \beta^k D_k Z
        - \alpha \kappa Z
        + \alpha W(\varphi) \left(
             D_i E^i
            - \mathcal{J}
        \right).
\end{align}

\subsection{Stress-energy tensor}\label{appsec:set}

The $3 + 1$ Einstein equations are written in terms of the following projections of the
stress-energy tensor (see, e.g.~\cite{Baumgarte:2010ndz}):
\begin{subequations}\label{eqn:stress-tensor-decomposition}
\begin{align}
    \rho
    &\equiv n^\alpha n^\beta T_{\alpha\beta} \\
    S_\mu
    &\equiv - \gamma_{\mu}^{\hphantom{\mu}\alpha} n^\beta T_{\alpha\beta} \\
    S_{\mu\nu}
    &\equiv \gamma_{\mu}^{\hphantom{\mu}\alpha} \gamma_{\nu}^{\hphantom{\nu}\beta} T_{\alpha\beta} \\
    S
    &\equiv \gamma^{\alpha\beta} S_{\alpha\beta}.
\end{align}
\end{subequations}
By construction, $S_\mu$ and $S_{\mu\nu}$ are spatial.

We write the total stress tensor as a sum of contributions from the scalar and gauge field, $T_{\mu
\nu} \equiv T_{\mu\nu}^\varphi + T_{\mu\nu}^A$.
The scalar-field stress-energy tensor,
\begin{align}
    T_{\mu\nu}^\varphi
    &= \nabla_\mu \varphi \nabla_\nu \varphi
        + g_{\mu\nu} \left(
            - \frac{1}{2} \partial_\alpha \varphi \partial^\alpha \varphi
            - V(\varphi)
        \right),
\end{align}
decomposes as
\begin{subequations}
\begin{align}
    \rho^\varphi
    &= \frac{1}{2} \Pi^2 + \frac{1}{2} D_i \varphi D^i \varphi + V(\varphi)
    \\
    S_\mu^\varphi
    &= - \Pi D_\mu \varphi
    \\
    S_{\mu\nu}^\varphi
    &= D_\mu \varphi D_\nu \varphi
        - \gamma_{\mu\nu} \left(
            - \frac{1}{2} \Pi^2
            + \frac{1}{2} D_i \varphi D^i \varphi
            + V(\varphi)
        \right)
    \\
    S^\varphi
    &= \frac{3}{2} \Pi^2
        - \frac{1}{2} D_i \varphi D^i \varphi
        - 3 V(\varphi).
\end{align}
\end{subequations}
The stress-energy tensor for the gauge field is
\begin{align}
    T_{\mu\nu}^A
    &= g^{\alpha\beta} F_{\mu\alpha} F_{\nu\beta}
        - g_{\mu\nu} \frac{1}{4} F_{\alpha\beta} F^{\alpha\beta},
\end{align}
and its $3 + 1$ components are
\begin{subequations}
\begin{align}
    \rho^A
    &= \frac{W(\varphi)}{2} \left( E_\alpha E^\alpha + B_\alpha B^\alpha \right)
    \\
    S_\mu^A
    &= W(\varphi) \epsilon_{\mu \alpha \rho} E^\alpha B^\rho
    \\
    S_{\mu\nu}^A
    &= W(\varphi) \left(
            - E_\mu E_\nu
            - B_\mu B_\nu
            + \frac{1}{2} \gamma_{\mu\nu}
                \left( E_\alpha E^\alpha + B_\alpha B^\alpha \right)
        \right)
    \\
    S^A
    &= \frac{W(\varphi)}{2} \left( E_\alpha E^\alpha + B_\alpha B^\alpha \right).
\end{align}
\end{subequations}

\section{Perturbation theory}\label{app:perttheory}

In this appendix, we detail perturbation theory in the BSSN formalism and describe how we set initial conditions in the BSSN code.

\subsection{Metric perturbations}
We define the perturbed metric
\begin{align}
    g_{\mu\nu}
    \equiv a(\tau)^2 \left( \eta_{\mu\nu} + h_{\mu\nu} \right)
\end{align}
where $h_{\mu \nu}$ is a small perturbation with scalar-vector-tensor decomposition
\begin{subequations}\label{eqn:metric-svt-decomposition}
\begin{align}
    h_{00}
    &= - E \\
    h_{i 0}
    &= \partial_i F + G_i  \\
    h_{i j}
    &= A \delta_{i j}
        + \partial_i \partial_j B
        + \partial_i C_j
        + \partial_j C_i
        + D_{ij}.
    \label{eqn:hij-pert-theory-def}
\end{align}
\end{subequations}
Here $C_i$ and $G_i$ are transverse vectors---namely, $\partial_i C_i = 0$ and
$\partial_i G_i = 0$---and $D_{ij}$ is transverse ($\partial_i D_{ik} = 0$) and traceless
($D_{ii} = 0$).

We expand the BSSN variables to linear order about a homogeneous and isotropic background spacetime as
\begin{subequations} \label{eq:bssnlin}
\begin{align}
    \alpha
    &= \alpha_0 + \delta \alpha
    \\
    \beta^i
    &= 0 + \delta \beta^i
    \\
    \phi
    &= \phi_0 + \delta \phi
    \\
    \bar{\gamma}_{i j}
    &= \delta_{i j} + \delta \gamma_{i j}
    \\
    K
    &= K_0 + \delta K
    \\
    \tilde{A}_{i j}
    &= 0 + a_{i j}.
\end{align}
\end{subequations}
Expanding the BSSN metric in these variables, we identify
\begin{align}\label{eqn:psi-0-alpha-0-scale-factor}
    e^{4\phi_0}
    = \alpha_0(\tau)^2
\end{align}
or
\begin{equation}
   \phi_0 = \frac{\ln \alpha_0}{2} ,
\end{equation}
and
\begin{align}
    \frac{\delta \alpha}{\alpha_0}
    &= E
    \\
    \beta_i
    &= a(\tau)^2 \left( \partial_i F + G_i \right)
    \\
    4 \delta \phi
    &= A
    \\
    \delta \gamma_{i j}
    &= \partial_i \partial_j B
        + \partial_i C_j
        + \partial_j C_i
        + D_{ij}.
\end{align}
We can linearize the equation of motion for $\bar{\gamma}_{i j}$ to obtain
\begin{align}
    \partial_t \bar{\gamma}_{m n}
    &= - 2 \alpha_0 a_{m n}
        + \delta_{m o} \partial_n \delta \beta^o
        + \delta_{o n} \partial_m \delta \beta^o
        - \frac{2}{3} \delta_{m n} \partial_o \delta \beta^o,
\end{align}
which allows us to identify gravitational waves (in the linearized theory) as
\begin{align}
    \partial_t D_{m n}
    &= - 2 \alpha_0 a_{m n}^\mathrm{T T}
\end{align}

The trace of the extrinsic curvature, $K$, expands to
\begin{align}
    K
    \equiv - \frac{6}{\alpha} \partial_t \phi
    &= - \frac{6}{\alpha_0}
        \left(
            1
            - \frac{\delta \alpha}{\alpha_0}
        \right)
        \left(
            \partial_t \phi_0
            + \partial_t \delta \phi
        \right)
    \\
    &= - \frac{6}{\alpha_0 }
        \left(
            \partial_t \phi_0
            + \partial_t \delta \phi
            - \partial_t \phi_0
            \frac{\delta \alpha}{\alpha_0}
        \right).
\end{align}
The background component is
\begin{align}
    K_0
    &= -\frac{6 }{\alpha_0 }\partial_t \phi_0,
\end{align}
which, using \cref{eqn:psi-0-alpha-0-scale-factor}, corresponds to the (cosmic-time) Hubble parameter $H = \mathcal{H}/a$ via
\begin{align}
    K_0
    = - \frac{6}{a} \frac{\partial_t \ln a}{2}
    = - 3 \frac{\partial_t a}{a^2}
    = - 3 H,
\end{align}
remembering that $\partial_t$ in the BSSN formalism corresponds to {\sl conformal time} for our particular gauge choices.

\subsection{Initial conditions}

The linearized tensor parts of $a_{i j}$ and $\delta \gamma_{i j}$, \cref{eq:bssnlin}, comprise genuine, propagating degrees of
freedom, so their initial condition is specified independent of constraints.
We therefore choose $D_{i j} = 0$ and $D_{i j}' = 0$ from \cref{eqn:metric-svt-decomposition} initially.
The remaining degrees of freedom are either gauge choices or set by constraints.
We take $\beta^i = \delta \beta^i = 0$ as an initial condition.
Correspondingly, we set $F = 0$ and $G_i = 0$ from \cref{eqn:metric-svt-decomposition}, amounting to the fixing of two vector and one scalar gauge
modes.
The initial conditions are simplest to solve for when taking $B = 0$ as the final gauge choice.

We define the short-hand
\begin{align}
    y(\mathbf{x})
    &= \frac{1}{\nabla^2} f(\mathbf{x})
    \equiv
        \int \frac{\ud^3 k}{(2 \pi)^3}
        \frac{e^{i \mathbf{k} \cdot \mathbf{x}}}{- k^2}
        \int \ud^3 y \,
            e^{- i \mathbf{k} \cdot \mathbf{y}}
            f(\mathbf{y})
\end{align}
to denote the solution to Poisson's equation, $\nabla^2 y(\mathbf{x}) = f(\mathbf{x})$.  The constraint on the vectors may be solved straightforwardly via
\begin{align}
    \frac{1}{2 \alpha_0} \partial_t C_m
    &= - \frac{1}{\Mpl^2} \frac{1}{\nabla^2} \mathcal{S}_m.
\end{align}
This sets the vector contribution $a_{m n}$ as
$- \partial_{(n} \partial_t C_{m)} / 2 \alpha_0$.  For the scalar degrees of freedom, with $B = 0$ one may directly solve the
scalar part of the momentum constraint as
\begin{align}
    \delta K
    &= - \frac{3}{2 \Mpl^2} \frac{1}{\nabla^2} \partial_m S_m.
\end{align}
With this solution, we can compute
\begin{align}
    \frac{\delta \psi}{\psi_0}
    &= \frac{\psi_0^4}{\nabla^2} \left(
            \frac{1}{6} K_0 \delta K
            - \frac{1}{4 \Mpl^2} \delta \rho
        \right).
\end{align}
Finally, Gauss's law expands to
\begin{align}
    \mathcal{J}
    = \partial_i E^i
    &= \partial_i \partial_t \mathcal{A}^i
        - \partial_i \partial^i \mathcal{A},
\end{align}
and, since we take $\partial_i \partial_t \mathcal{A}^i = 0$ initially, we satisfy this constraint by
setting
\begin{align}
    \mathcal{A}
    &= - \frac{1}{\nabla^2} \mathcal{J}.
\end{align}

\section{Robustness checks}\label{app:robustness}

\begin{figure}[t!]
    \centering
    \includegraphics[width=0.67\textwidth]{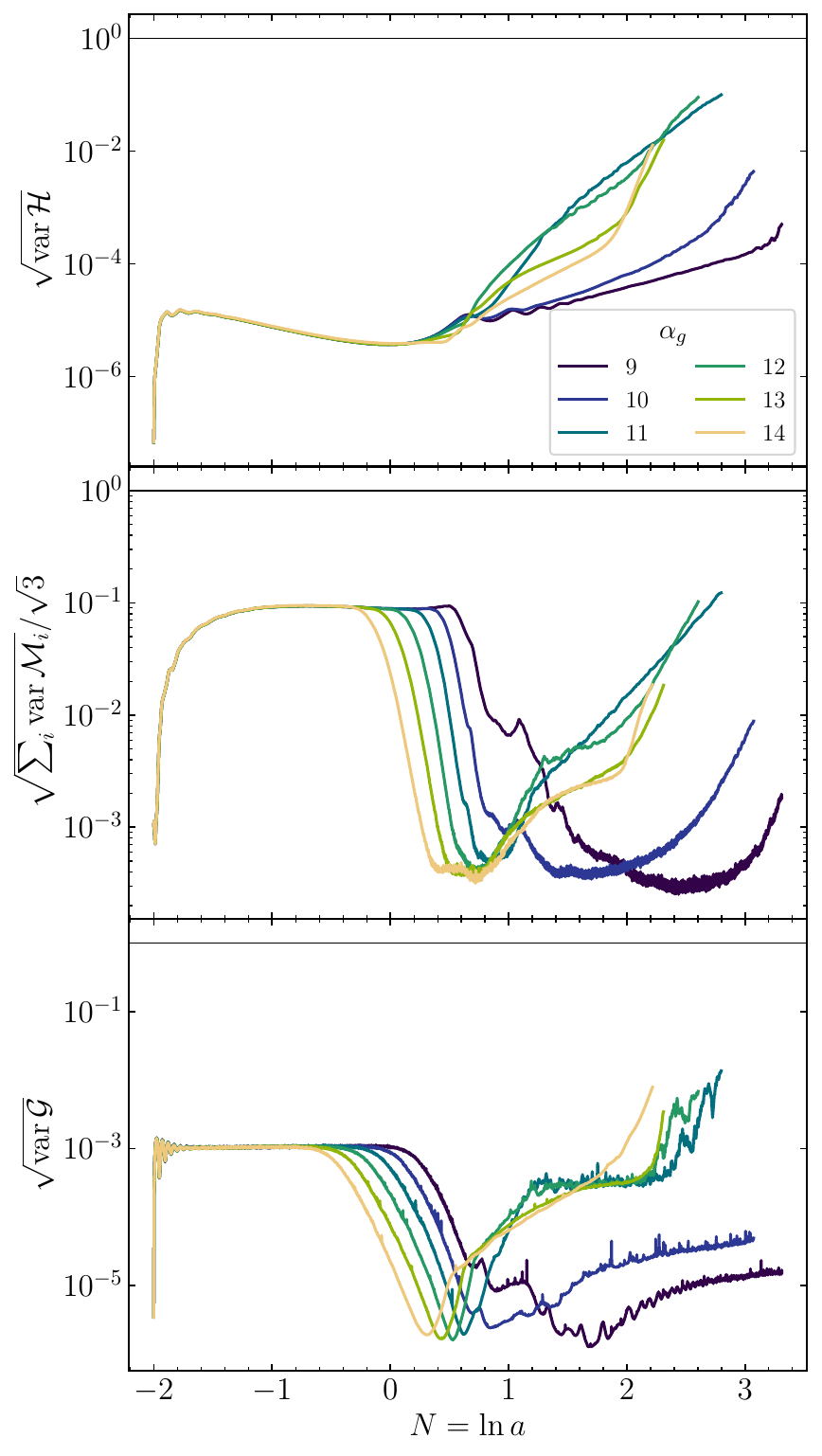}
    \caption{
        Statistics of violations of the Hamiltonian [$\mathcal{H}$, \cref{eq:hamconst}], momentum  [$\mathcal{M}_i$, \cref{eq:momconst}], and Gauss [$\mathcal{G}$, \cref{eq:GaussLaw}] constraints.  Note that we in the main text we only present results up to $N = 2$.
    }
    \label{fig:contraints}
\end{figure}

For the six BSSN simulations presented here, we calculate the violation of the Hamiltonian and momentum constraints, \cref{eq:hamconst,eq:momconst}, and of Gauss' Law, \cref{eq:GaussLaw}, to ensure that we stay on the solution surface of the problem.  In \cref{fig:contraints} we plot these constraints vs $N \equiv \ln a$ to show that they are bounded throughout the simulation.
For the results presented in this work, we take a conservative cutoff of $N = 2$ as a final time.  This allows us to consider all of the BSSN simulations in regime where the constraints are small.
In addition, the results in \cref{sec:results} are well reproduced by simulations with $256^3$
rather than $384^3$ grid points; the latter is the largest grid that is computationally tractable
with our resources.
We choose this largest grid for our main results in order to maximize the range in scales over which
the BSSN results may be reliably compared to the FLRW results, e.g., in
\cref{fig:delta-spectra-bssn-vs-flrw}.

For FLRW simulations, the first Friedmann equation, $H^2 = \rho / 3 \Mpl^2$, serves as a constraint that measures
the analog of energy conservation in flat space; the simulations satisfy it to one part in $10^{4}$ or better.
The gauge constraints are likewise satisfied to one part in $10^{3} \div 10^{2}$ (see Refs.~\cite{Adshead:2015pva,Adshead:2016iae,Adshead:2018doq,Adshead:2019lbr} for further discussion of the robustness of these methods).
The FLRW results are all consistent with those of lower-resolution simulations presented in Refs.~\cite{Adshead:2018doq,Adshead:2019lbr,Adshead:2019igv}.

\bibliographystyle{JHEP}
\bibliography{NRGaugePreheat}

\end{document}